\documentclass[twocolumn,preprintnumbers,amssymb,amsmath,superscriptaddress,letterpaper]{revtex4}[10pt]

\usepackage{graphicx}
\usepackage{dcolumn}
\usepackage{bm}
\usepackage{natbib}
\usepackage{pstricks}
\usepackage{epsfig}

\newcommand{\be}{\begin{equation}}
\newcommand{\ee}{\end{equation}}
\newcommand{\bea}{\begin{eqnarray}}
\newcommand{\eea}{\end{eqnarray}}

\newcommand{\gsim}{ \mathop{}_{\textstyle \sim}^{\textstyle >} }
\newcommand{\lsim}{ \mathop{}_{\textstyle \sim}^{\textstyle <}}

\begin{document}

\title{Searching for High Energy Neutrino counterpart signals; The case of the Fermi Bubbles signal and of Dark Matter annihilation in the inner Galaxy}

\author{Ilias Cholis}
\email{ilias.cholis@sissa.it}
\affiliation{SISSA, Via Bonomea, 265, 34136 Trieste, Italy}
\affiliation{INFN, Sezione di Trieste, Via Bonomea 265, 34136 Trieste, Italy}

\date{\today}

\begin{abstract}

The recent uncovering of the \textit{Fermi} Bubbles/haze in the 
\textit{Fermi} $\gamma$-ray data has generated theoretical work 
to explain such a signal of hard $\gamma$-rays in combination with 
the \textit{WMAP} haze signal. Many of these theoretical models can have
distinctively different implications with regards to the production
of high energy neutrinos. We discuss the neutrino signals from
different models proposed for the explanation of the \textit{Fermi} 
Bubbles/haze, more explicitly, from  
Dark Matter annihilation in the galactic halo with conditions of 
preferential CR diffusion, from recent AGN jet activity, from periodic 
diffusive shock acceleration, from stochastic 2nd order Fermi 
acceleration and from long time-scale star formation in the galactic 
center in combination with strong galactic winds.
We find that some of these models will be probed by the IceCube 
DeepCore detector. Moreover, with a km$^3$ telescope located at the 
north hemisphere, we will be able to discriminate between the 
hadronic, leptonic and the DM models. Additionally using the reconstructed 
neutrino spectra we will probe annihilation of TeV scale dark matter 
towards the galactic center.

\end{abstract}
\maketitle

\section{Introduction}
\label{sec:intro}

Using the first year of \textit{Fermi} LAT $\gamma$-ray full sky data, 
\cite{Dobler:2009xz} revealed the presence of a diffuse component 
towards the galactic center that extends up to $50^{\circ}$ in latitude.
That component has a spectrum significantly harder than elsewhere in 
the Galaxy \cite{Dobler:2009xz, Su:2010qj, Dobler:2011mk} and a 
morphology elongated in latitude to longitude, that depending on the 
exact template analysis used for its extraction from the full sky data, 
is either well defined within 2 bubbles with distinct edges both at high 
and low latitudes \cite{Su:2010qj}
 known as the "\textit{Fermi} Bubbles", or is slightly more diffuse 
(the "\textit{Fermi} haze") with a latitude to longitude axis ratio 
of $\simeq 2$ \cite{Dobler:2011mk} but still confined within 
$\mid l \mid \lsim 20^{\circ}$ and $\mid b \mid \lsim 50^{\circ}$ and with 
the "edges" seen only at the higher latitudes \cite{Dobler:2011mk}.  

That signal of hard $\gamma$-rays together with the \textit{WMAP} haze
\cite{Finkbeiner:2003im, Dobler:2007wv} may indicate the presence of a hard 
component of Cosmic Ray (CR) 
electrons, that through their up-scattering of low energy photons produce
the hard spectrum of (Inverse Compton) $\gamma$-rays and through their
synchrotron radiation the hard observed spectrum at microwaves. Various 
authors have suggested mechanisms for the origin of these CR electrons.
Among them, possible scenarios include recent (1-3 Myr ago) AGN jet 
activity in the galactic center \cite{Guo:2011eg, Guo:2011ip}; TeV scale 
Dark Matter (DM) annihilating to leptons \cite{Cholis:2009va, Dobler:2011mk}, 
within conditions of preferential diffusion perpendicular to the plane 
\cite{Dobler:2011mk}; stochastic 2nd order Fermi acceleration by large
scale turbulence in magneto-sonic waves \cite{Mertsch:2011es}, or
periodic injection of hot plasma causing diffusive shock acceleration 
(1st order Fermi acc.) in the halo \cite{Cheng:2011xd}.

Alternatively, CR protons associated with long time scale ($\sim$ Gyr) star 
formation in the galactic center, transferred by strong winds into 
the \textit{Fermi} Bubbles region have been suggested by \cite{Crocker:2010dg}.
Finally, a combination of DM and millisecond pulsars in the galactic halo 
adding up to the signals at $\gamma$-rays and microwaves has been discussed in
\cite{Malyshev:2010xc}. 

The detection or lack of high energy neutrinos from km$^3$ neutrino telescopes
could help discriminate between the leptonic \cite{Guo:2011eg, Guo:2011ip, Dobler:2011mk, Mertsch:2011es, Cheng:2011xd} and the hadronic \cite{Crocker:2010dg}
scenarios, since the leptonic scenarios would not produce any, or a few and up to 
the TeV scale neutrinos while the hadronic explanation of \cite{Crocker:2010dg} 
would produce abundant neutrinos up to the PeV scale \cite{Vissani:2011ea, Gaisser:2012ru} 
(see also discusion in \cite{Lunardini:2011br}).

In addition to discriminating among different models for the 
\textit{Fermi} haze/Bubbles via searching for their neutrino counterpart, 
such searches can be used as an other channel of indirect DM searches.
 
DM composes approximately $85 \%$ of the matter density of the universe,
yet its particle physics properties still remain unknown.
Measurements of cosmic rays (CR)
\cite{Boezio:2008mp, Adriani:2008zr, Adriani:2008zq,Collaboration:2008aaa, Aharonian:2009ah, Abdo:2009zk, ATIClatest}
have generated new model building \cite{ArkaniHamed:2008qn, Cholis:2008qq, Harnik:2008uu, Fox:2008kb,Grajek:2008pg, Pospelov:2008jd, MarchRussell:2008tu, Cirelli:2009uv, Shepherd:2009sa, Phalen:2009xw, Hooper:2009fj, Cholis:2009va, Hooper:2009gm}
and have helped place new constraints on dark matter properties
\cite{Cirelli:2008pk, Donato:2008jk, Cholis:2010xb, Evoli:2011id, Bergstrom:2009fa}.
Since many of these models and constraints are placed at the TeV mass scale
and suggest/refer to enhanced (boosted) annihilation cross-sections with
hard spectra for the Standard Model (SM) particle annihilation products, neutrino signals
from the galactic center (GC) \cite{Erkoca:2010qx} from the Sun \cite{Mandal:2009yk}
or the Earth \cite{Albuquerque:2011ma} can be considered of interest.

IceCube in the South Pole \cite{Abbasi:2010rd, Collaboration:2011ym} has 
already presented some early (pre-DeepCore) results \cite{IceCube:2011ae, Heros:2010ss, Abbasi:2011eq} 
and is expected with its DeepCore update to better probe the region that is sensitive for 
DM searches and has  $\lsim$100 GeV in neutrino energy.
ANTARES located in the Mediterranean is also collecting data 
\cite{AdrianMartinez:2011uh, Bogazzi:2011zza, Heijboer:2011zz} and due to
its location probes significantly better the GC than the IceCube, but because 
of its small size has still low number of statistics. Finally a future 
km$^3$ telescope located in the Mediterranean as KM3NeT 
\footnote{http://www.km3net.org/home.php ,\\ 
http://www.km3net.org/public.php}
\cite{Tsirigotis:2012nr, Leisos:2012zi} will combine the virtues of IceCube 
and ANTARES providing a good neutrino telescope for searches of DM 
annihilation towards the GC.

In this paper, in section~\ref{sec:LDMA} we will discuss the neutrino signals
of various DM models that could explain the \textit{Fermi} haze and 
\textit{WMAP} haze signals as presented in \cite{Dobler:2011mk}, for both IceCube 
and a future km$^3$ telescope located in the northern hemisphere (using the 
Mediterranean as the Earth's latitude of reference), presenting also 
search strategies for those signals. We use simulated performance information 
published for the KM3NeT \cite{Tsirigotis:2012nr, Leisos:2012zi}).
We will extend in section~\ref{sec:ODMA} the discussion of searching for signals in 
neutrinos towards the GC from more generic DM models.
In section~\ref{sec:HSFB} we will describe the neutrino predictions of the hadronic 
model of \cite{Crocker:2010dg} for the \textit{Fermi} Bubbles that if true should
soon be seen. In section~\ref{sec:alternate_models} we discuss why we don't 
expect any significant signal in neutrinos from the non-DM leptonic models of 
\cite{Guo:2011eg, Guo:2011ip, Mertsch:2011es, Cheng:2011xd} presented for the Bubbles 
and conclude in section~\ref{sec:conclusions}.

In this work we will discuss the upward going $\nu_{\mu}$s flux
expected to be measured at the ongoing and future telescopes. 
For a comparison of the upward going muon events rate and the rate of fully contained muons
produced by nuetrinos inside the detector see discussion in \cite{Erkoca:2010vk}.

\section{eXciting Dark Matter annihilation, connecting to the \textit{Fermi} haze and the \textit{WMAP} haze}
\label{sec:LDMA}

In the context of DM annihilating to leptons, \cite{Dobler:2011mk} invoked preferential 
(anisotropic) diffusion of CR electrons perpendicular to the galactic disk due to 
ordered magnetic fields in the same direction. Such conditions could explain the 
\textit{Fermi} haze
spectrum and morphology in combination with the \textit{WMAP} haze spectrum and angular
profile (see Figs. 7 and 8 of \cite{Dobler:2011mk}). 

In that case, eXciting Dark Matter (XDM) \cite{Finkbeiner:2007kk} was considered where 
DM particles with mass $m_{\chi}=1.2$ TeV annihilate into a pair of scalar bosons $\phi$ that
then decay to a pair of $e^{\pm}$ due to kinematic suppression for the case of 
$m_{\phi} < 2 m_{\mu}$. In such a case the energy released by the DM annihilation 
goes to $e^{\pm}$ that have a hard enough spectrum to explain the haze signals
\cite{Dobler:2011mk, Cholis:2009va}.  

Also in an XDM scenario as has been shown in \cite{ArkaniHamed:2008qn}, the 
annihilation cross-section can be enhanced up to $O(10^3)$ to motivate the 
necessary boost (BF) on the annihilation rates that have been invoked to
explain the CR positron fraction and $e^{-}+e^{+}$ flux excesses 
\cite{Cholis:2008qq, Cholis:2008wq, Finkbeiner:2010sm}. Thus the magnitude of 
the annihilation rate needed to explain the \textit{Fermi} and \textit{WMAP}
haze signals (BF$\simeq 30$), is naturally explained within the context of 
an XDM annihilating to final state SM leptons. Finally, since from Cold Dark 
Matter (CDM) cosmological 
simulations \cite{Diemand:2008in, Kuhlen:2008qj, Springel:2008cc} the DM halo 
profiles are typically triaxial, a prolate DM profile has been used in 
\cite{Dobler:2011mk}, with its axis perpendicular to the galactic disk. Observations
of the spatial distribution of Milky-Way satellites suggest a prolated DM halo with its
major axis perpendicular to the stellar disk \cite{zentner:2005ad}, in agreement with suggestions
by hydrodynamic simulations of galaxies with stellar disks \cite{Bailin:2005xq}. 

For the case when the SM leptons from the $\phi$ decay are $e^{\pm}$ 
($m_{\phi} < 2 m_{\mu}$), no neutrinos are produced thus the presence of a 
possible ``Neutrino haze'' is excluded. Alternatively, for $m_{\phi} > 2 m_{\mu}$ 
the decay to $\mu^{\pm}$ is allowed and a neutrino haze can exist. To maximize   
the possible neutrino haze signal and also to explain the \textit{Fermi} and 
\textit{WMAP} signals we will consider the case where $m_{\chi} = 2.5$ TeV 
particles annihilate to a pair of $\phi$s that decay with a BR=1 to $\mu^{\pm}$ 
\footnote{The BR to muons for $2 m_{\mu} < m_{\phi} < 2 m_{\pi}$ can be 
dominant for a \textit{scalar} $\phi$.}. The mass of 2.5 TeV is chosen to 
produce -after the muons decay- $e^{\pm}$ which during propagation will give 
similar synchrotron and IC signals. The necessary enhancement in the annihilation
rate is BF$\simeq$150 to produce the same total injected energy in high energy  
$e^{\pm}$ (see for instance Fig. 6 \& 7 of \cite{Cholis:2008wq}). 

Since neutrinos do not diffuse or loose energy, the neutrino signal from the 
DM halo will be identical to the annihilation rate profile 
$\sim \int \langle \sigma v \rangle \rho_{DM}^{2} dl d\Omega$; with $l$ being the 
line of sight, $d\Omega$ the angle of observation, $\rho_{DM}$ the DM density and
$\langle \sigma v \rangle$ the velocity averaged annihilation cross-section
(more accurately written as $\langle \sigma \mid v \mid \rangle$).
Since in the Sommerfeld enhancement case, the $\langle \sigma v \rangle$ depends 
on the velocity dispersion 
\cite{SommerfeldRef, ArkaniHamed:2008qn, Lattanzi:2008qa, Hisano:2006nn}, 
it also may have a profile within the main halo (see for instance \cite{Robertson:2009bh}). That is ignoring effects of 
substructure that may make the position dependence of the averaged (after integration)
$\langle \sigma v \rangle$ over the galaxy even more evident 
\cite{Slatyer:2011kg}.  

\begin{figure*}[t]
\hspace{-0.8cm}
\includegraphics[width=3.60in,angle=0]{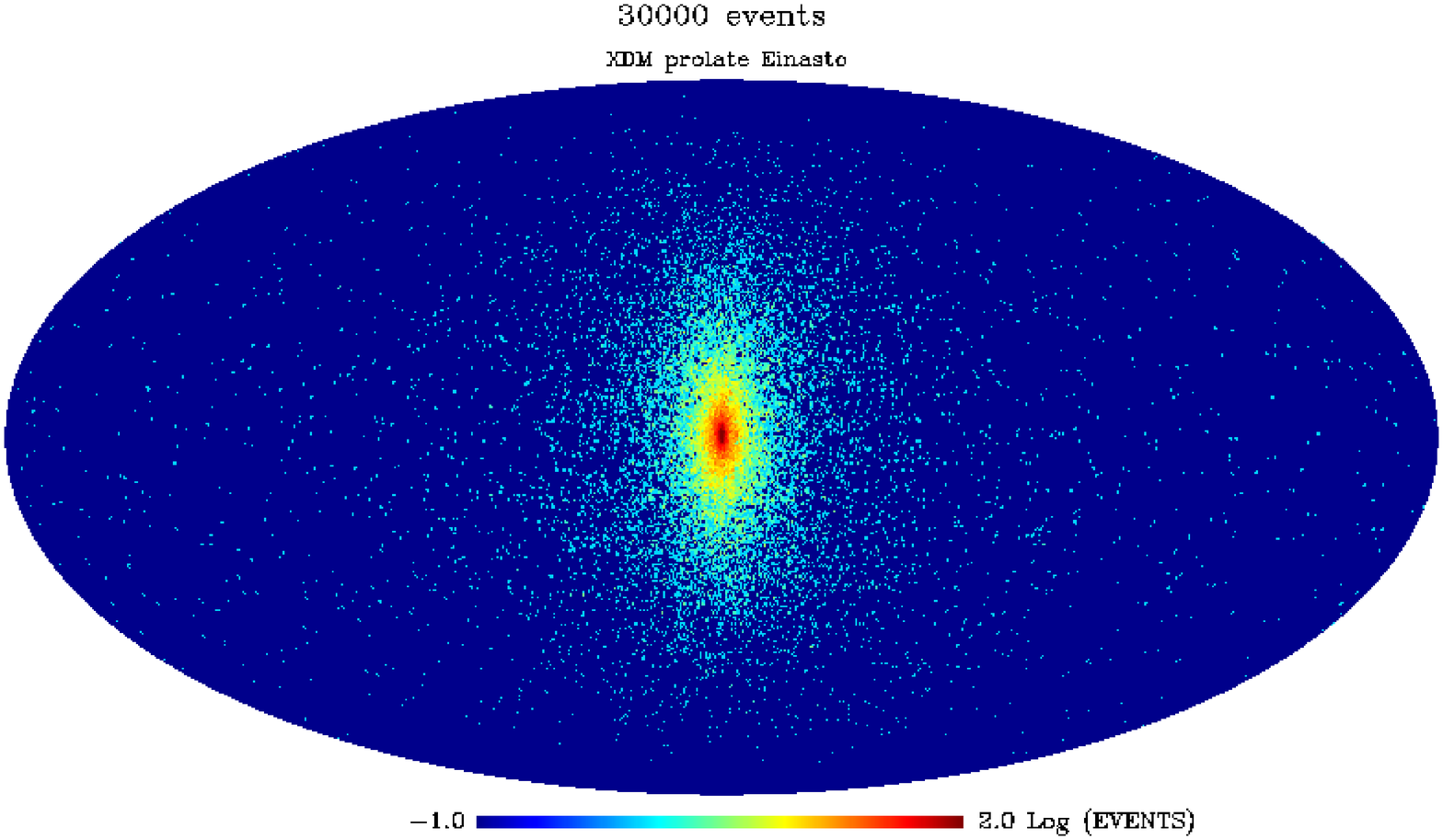}
\hspace{-0.2cm}
\includegraphics[width=3.60in,angle=0]{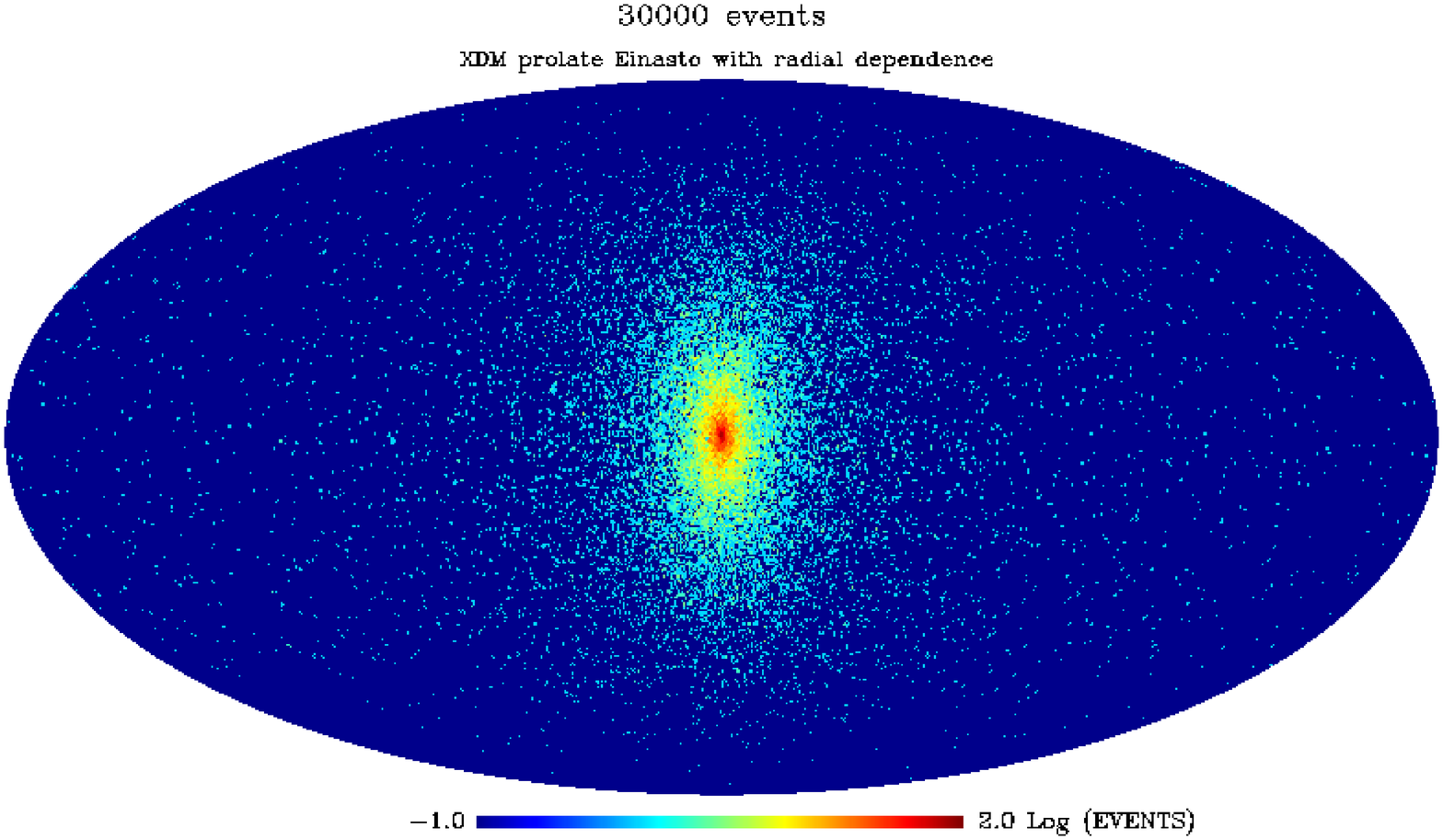}
\caption{$3\times 10^{4}$ simulated $\nu$ events, from XDM to $\mu^{\pm}$ scenario. 
\textit{Left}:Prolate Einasto profile with homogeneous enhancement, \textit{right}: including
a $\propto r^{1/4}$ in the annihilation cross-section. The latter case is less prolate in its
morphology. We present the neutrino maps in Mollweide projection using HEALPix \cite{Gorski:2004by}.}
\label{fig:DMmaps}
\end{figure*}

In Fig.~\ref{fig:DMmaps} we show the case of $3\times 10^{4}$ neutrino
simulated events for a prolate DM Einasto profile described by:
\begin{equation}
  \rho(z,R)=\rho_{0} exp \left[\frac{2}{\alpha} 
\frac{R_{\odot}^{\alpha}}{R_{c}^{\alpha}}\right] 
exp\left[-\frac{2}{\alpha}\left(\frac{R^{2}}{R_{c}^{2}} + 
\frac{z^{2}}{Z_{c}^{2}}\right)^{\alpha/2}\right]
  \label{eq:ProlateEin}
\end{equation}
with $\rho_{0} = 0.4$ GeV cm$^{-3}$ the local DM density \cite{Catena:2009mf, Salucci:2010qr},
$\alpha = 0.17$, $Z_{c}/R_{c} =2$ and $Z_{c}=27$ kpc
(giving a total amount of DM within the inner $\sim 100$ kpc as is for a spherically 
symmetric case of $Z_{c} = R_{c} =25$ kpc) \cite{Merritt:2005xc}.

The cases of a homogeneous $\langle \sigma v \rangle$ Fig.~\ref{fig:DMmaps} (left)
and of $\langle \sigma v \rangle \propto r^{1/4}$ in Fig.~\ref{fig:DMmaps} (right),
(with $r$ the galactocentric distance) are shown. The latter dependence can be the 
case for the Sommerfeld models \cite{Cholis:2009va}, as cosmological simulations 
with DM and baryons have suggested a profile of increasing velocity dispersion 
towards the GC (compared to the local values) and up to the inner 1 kpc
\cite{RomanoDiaz:2009yq, Pedrosa:2009rw, Tissera:2009cm, Parry:2011iz}.   
The profile with the $\propto r^{1/4}$ cross section is slightly more diffuse 
and less elongated. Since the exact DM profile in the inner $5^{\circ}$ of 
the galactic center is very uncertain we will derive our conclusions ignoring
 that part of the DM halo. Additionally, TeV neutrino sources either 
point or diffuse (from inelastic collisions of CR nuclei with dense ISM gas)
are concentrated along the galactic disk. The thin galactic disk 
where continuous star formation takes place, has a characteristic (for 
exponentially decreasing) scale high of 0.3 kpc\cite{2008gady.book.....B}. 
This scale hight is indicative of the majority of the TeV neutrino sources
\footnote{For IceCube and KM3NeT the expected angular resolution of the 
reconstructed events is expected to be a few degrees 
\cite{Abbasi:2010rd, Tsirigotis:2012nr, Lenis:2012uw}.}.
The HI ISM gas has also a scale hight that towards the GC is $\simeq 0.15$ 
kpc and increasing up to 0.3 kpc at the solar ring 
\cite{2008A&A...487..951K, 2007A&A...469..511K}; while the H2 gas a scale
hight of 0.1 kpc \cite{1988ApJ...324..248B, 2006PASJ...58..847N}. These scale 
hights are indicative for the galactic
diffuse TeV neutrino flux\footnote{For profiles in diffuse $\gamma$-rays
of the equivalent "pi0" component see\cite{FermiLAT:2012aa}.} (see also 
work of \cite{Evoli:2007iy}).  
Thus  we will avoid the entire inner $5^{\circ}$ in $\mid b \mid$.   

The neutrino flux at Earth due to DM annihilation from an 
angle $d\Omega$ \textit{ignoring oscillation} ($\phi^{0}_{\nu^{i}}$) is 
described by:\footnote{for a Majorana particle}

\begin{equation}
\frac{d\phi^{0}_{\nu^{i}}}{dE_{\nu^{i}}} = \int d\Omega \int_{l.o.s.} 
d \ell ({\theta}) \frac{\rho_{DM}^{2}\langle\sigma v \rangle(\ell,\theta)}
{8\pi m_{\chi}^{2}} \frac{dN_{\nu^{i}}}{dE_{\nu^{i}}},
 \label{eq:neutrinoflux0}
\end{equation}
where we have left $\langle\sigma v \rangle(\ell,\theta)$ to depend on the
position in the Galaxy for the most generic case. A boost factor is absorbed 
in either $\rho_{DM}^{2}$ or/and the $\langle\sigma v \rangle$. The 
$\frac{dN_{\nu^{i}}}{dE_{\nu^{i}}}$ is the neutrino spectrum of the species 
$\nu_{i}$. The multiplicity $M^{i}$ of $\nu_{i}$s per annihilation event is 
absorbed in $\frac{dN_{\nu^{i}}}{dE_{\nu^{i}}}$ giving:
\begin{equation}
\int^{m_{\chi}}_{0} \frac{dN_{\nu^{i}}}{dE_{\nu^{i}}} dE = M^{i}.
 \label{eq:Multiplicity}
\end{equation} 

In this work we  discuss only the upward going $\nu_{\mu}$s flux. 
The $A_{eff}$ of $\nu_{e}$
upward for both the IceCube DeepCore (not optimally placed for the GC searches)
\cite{Collaboration:2011ym} and the KM3NeT \cite{PrivateCom}, is smaller 
for $\nu_{e}$s by at least a factor of 2 at all energies of interest. For 
simplicity we are going to ignore their contribution.

The observed $\nu_{\mu}$ flux at Earth after oscillations is given by 
\cite{Abbasi:2011eq, Barenboim:2003jm, Murase:2007yt}:
\begin{eqnarray}
\phi_{\nu_{\mu}} &\simeq& \frac{1}{2} \left(\phi^{0}_{\nu_{\mu}} + 
\phi^{0}_{\nu_{\tau}} \right) + \frac{1}{8}s_{2} \; \; 
\textrm{with} \nonumber\\
s_{2} &=& sin^{2}2\Theta_{12} \left(2\phi^{0}_{\nu_{e}} - 
\phi^{0}_{\nu_{\mu}} - \phi^{0}_{\nu_{\tau}} \right) \; \; \textrm{and} \\
sin^{2}2\Theta_{12} &=& 0.86, \nonumber
 \label{eq:Oscilation}
\end{eqnarray} 
where $\phi^{0}_{\nu_{i}}$ is the flux at injection of flavor species
$\nu_{i}$.

For specific experiments one has to include the strong dependence
of the telescopes effective area with angle and energy. Within an
angle $d\Omega$ and an energy range $E$ - $E + \Delta E$ the total 
number of upward going $\nu_{\mu} + \overline{\nu}_{\mu}$ events is
\footnote{The charged current and 
neutral current neutrino nucleon cross sections are different for 
$\nu_{\mu}$ and $\overline{\nu}_{\mu}$\cite{Strumia:2006db, Erkoca:2010qx}.
Here we use averaged values of $A_{\nu_{\mu}}$, $A_{\overline{\nu}_{\mu}}$.
For the DM annihilation cases that we study the $\nu_{\mu}$ and 
$\overline{\nu}_{\mu}$ fluxes arriving at Earth are equal.}:  

\begin{eqnarray}
  N_{\nu_{\mu},\overline{\nu}_{\mu}}(E,d\Omega)&=& \int^{E'+\Delta E'}_{E'} dE' \int d\Omega 
\int_{l.o.s.} d\ell(\theta) 
\frac{\rho_{DM}^{2}\langle\sigma v \rangle}
{8\pi m_{\chi}^{2}} \nonumber \\
 & & A_{eff_{{\nu_{\mu}, \overline{\nu}_{\mu}}}}(E', \theta)
\frac{dN^{osc.}_{\nu_{\mu}, \overline{\nu}_{\mu}}}             
{dE'_{\nu_{\mu}, \overline{\nu}_{\mu}}} \; ,
  \label{eq:EventsNumber}
\end{eqnarray}
where,
\begin{eqnarray}
\frac{dN^{osc.}_{\nu_{\mu}, \overline{\nu}_{\mu}}}
{dE'_{\nu_{\mu}, \overline{\nu}_{\mu}}} &=& 
\frac{1}{2}\left(\frac{dN_{\nu_{\mu}, \overline{\nu}_{\mu}}}
{dE'_{\nu_{\mu}, \overline{\nu}_{\mu}}} + 
\frac{dN_{\nu_{\tau}, \overline{\nu}_{\tau}}}
{dE'_{\nu_{\tau}, \overline{\nu}_{\tau}}} \right) \\
&+& \frac{1}{8}0.86 \left(2 
\frac{dN_{\nu_{e}, \overline{\nu}_{e}}}
{dE'_{\nu_{e}, \overline{\nu}_{e}}} - 
\frac{dN_{\nu_{\mu}, \overline{\nu}_{\mu}}}
{dE'_{\nu_{\mu}, \overline{\nu}_{\mu}}} 
- \frac{dN_{\nu_{\tau}, \overline{\nu}_{\tau}}}
{dE'_{\nu_{\tau}, \overline{\nu}_{\tau}}} \right). \nonumber
 \label{eq:Oscilated}
\end{eqnarray}

For the $\frac{dN_{\nu^{i}}}{dE_{\nu^{i}}}$ originating from the 2.5 TeV 
XDM to muons case, the injection spectra of $\nu_{\mu}$, $\overline{\nu}_{\mu}$, 
$\nu_{e}$, $\overline{\nu}_{e}$ (there are no $\nu_{\tau}$s) are practically 
identical to those of the injected $e^{\pm}$ given in appendix A of 
\cite{Cholis:2008vb}. Per annihilation event there are 2 neutrinos 
(and 2 antineutrinos) for each flavor.

Having excluded the $\mid b \mid < 5^{\circ}$ region, the basic remaining 
background is that of the atmospheric upward neutrinos. The atmospheric 
background flux is isotropic after averaging for the many different directions 
of the neutrino telescopes axis within long timescales. 
For the atmospheric  $\nu_{\mu}$ and $\overline{\nu}_{\mu}$ spectra 
and fluxes above 10 GeV and up to 10 TeV we used the tables of Appendix B 
of \cite{Honda:2006qj}, extrapolating to higher energies with a spectral 
power law of 3.7 for the differential spectrum \cite{Erkoca:2010qx}.

\begin{figure*}[t]                                                          
\vspace{-0.3cm}
\hspace{-0.8cm}
\includegraphics[width=3.60in,angle=0]{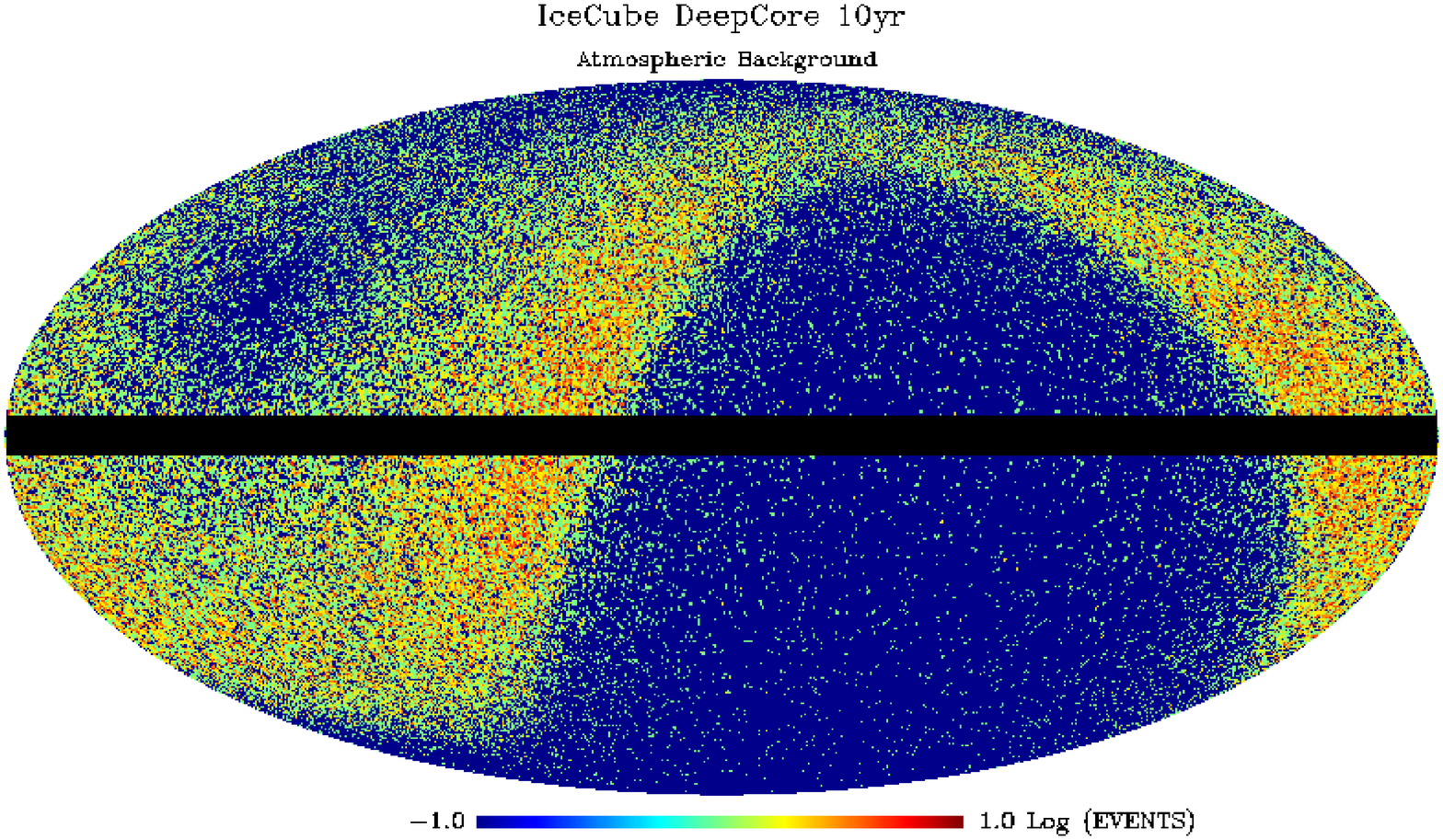}
\hspace{-0.2cm}
\includegraphics[width=3.60in,angle=0]{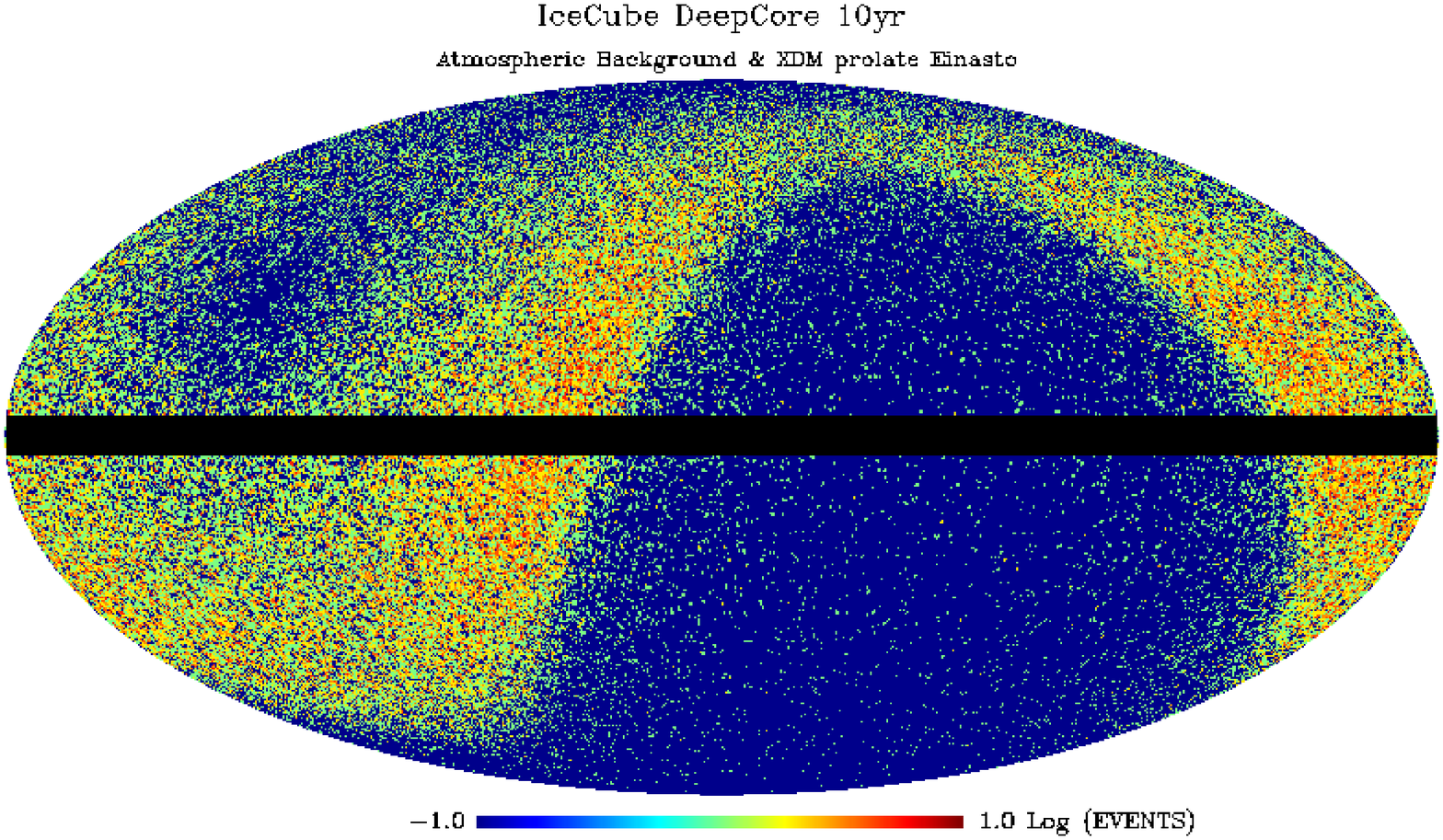} \\
\vspace{-0.3cm}
\includegraphics[width=3.60in,angle=0]{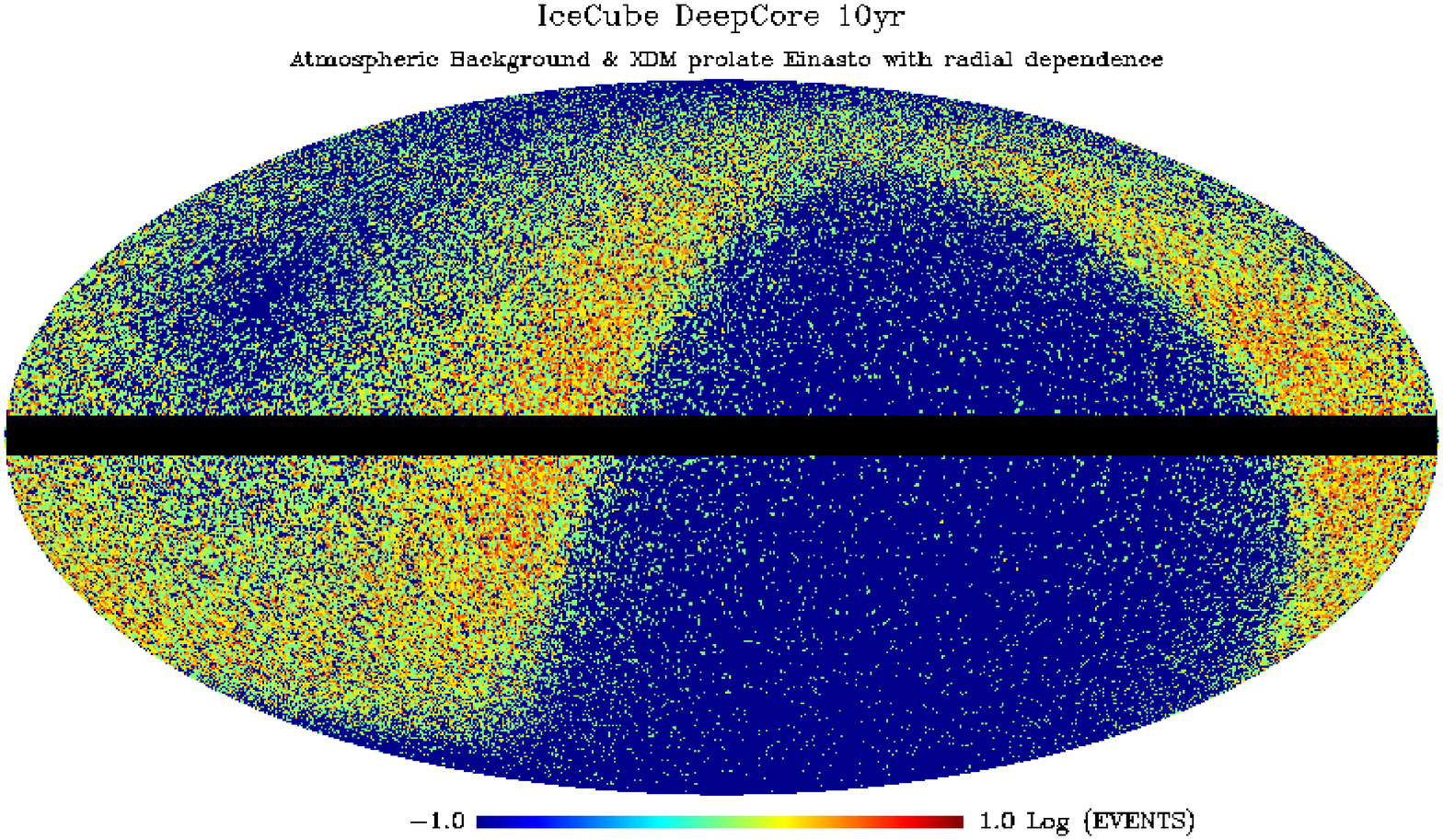} \\
\hspace{1.7cm}
\includegraphics[width=3.30in,angle=0]{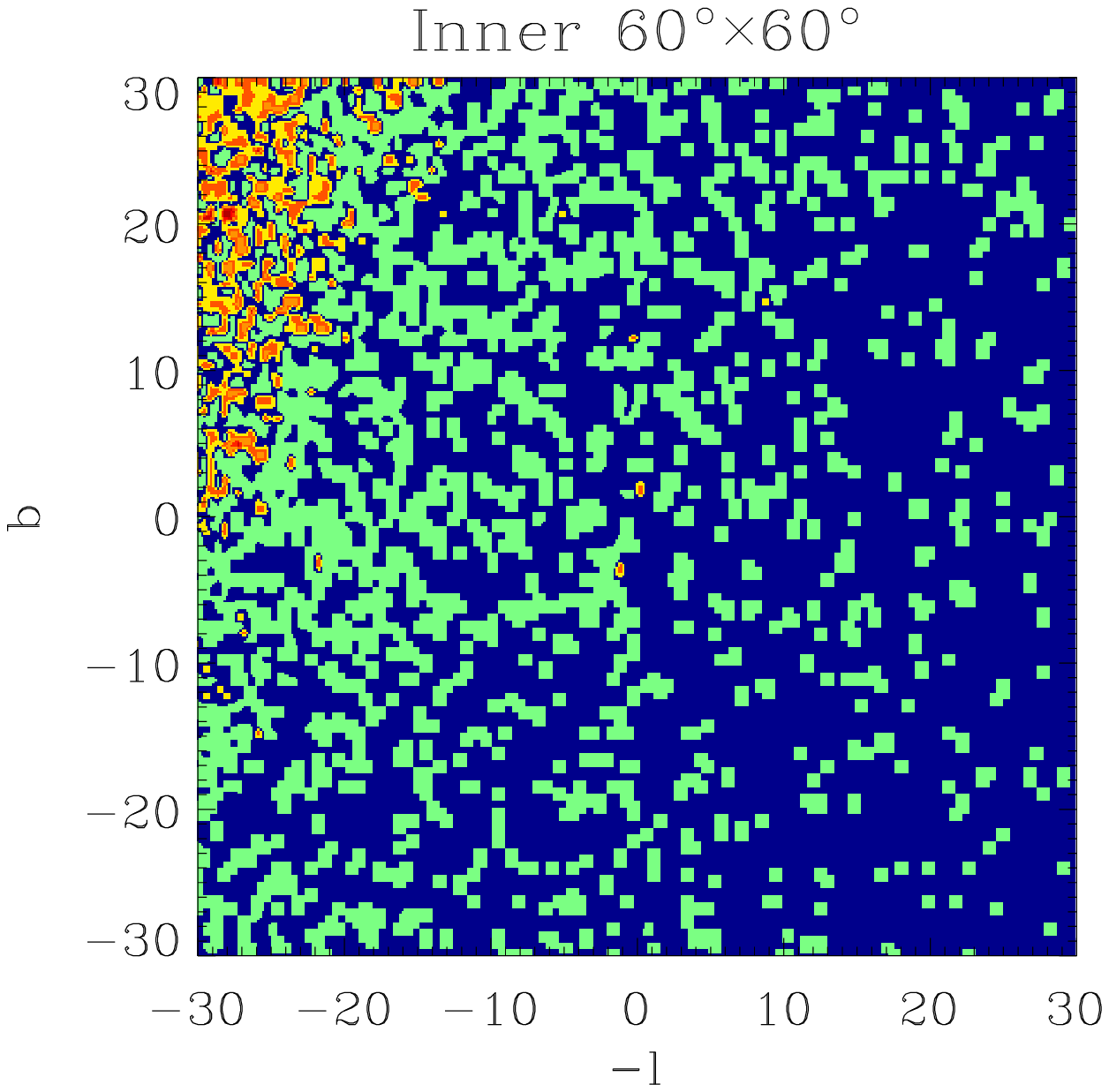}
\hspace{-1.0cm}
\includegraphics[width=3.30in,angle=0]{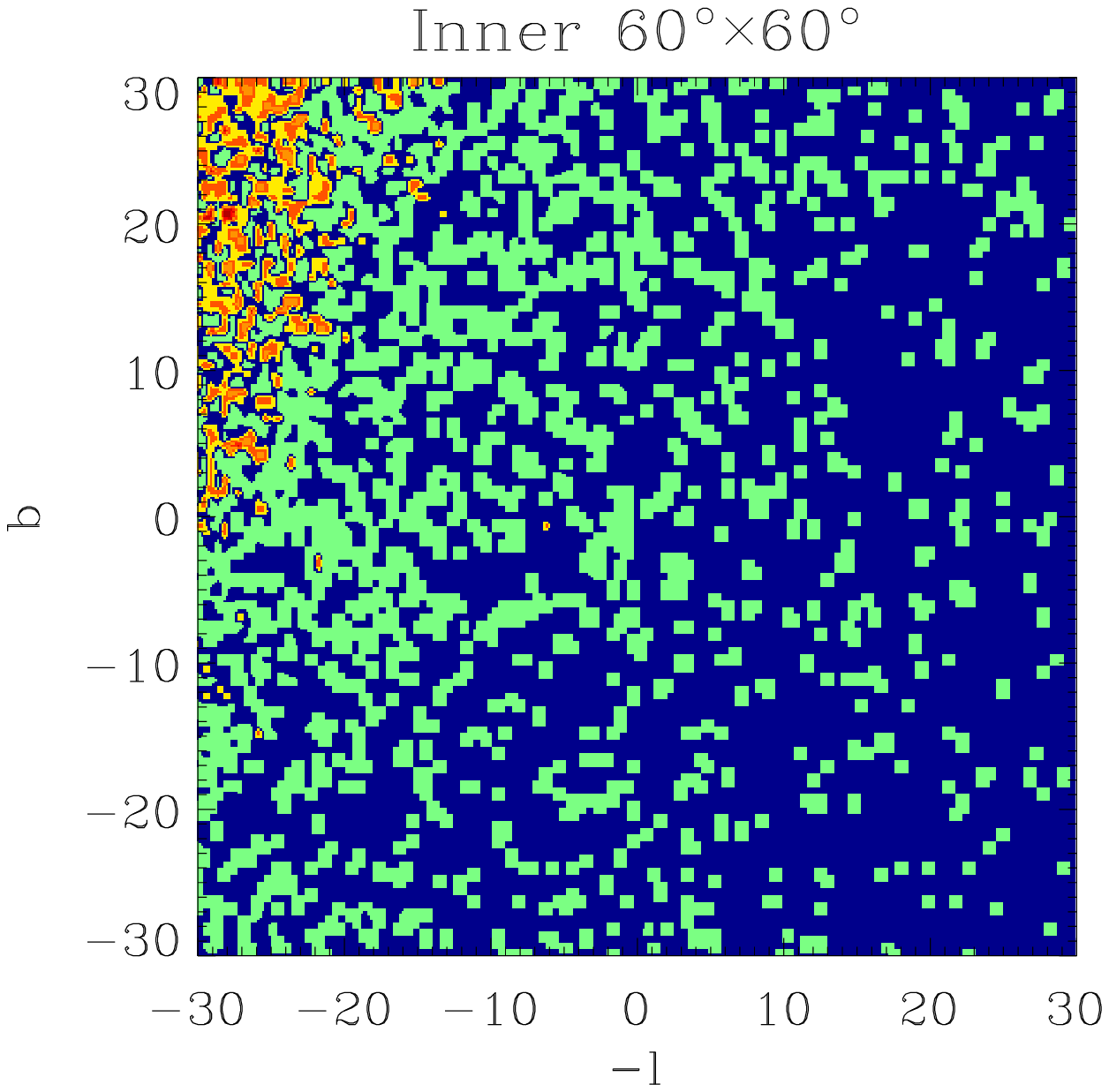}
\caption{IceCube DeepCore $\nu_{\mu} + \overline{\nu}_{\mu}$ events 
after 10 years of data collected with ``online filter'' 
\cite{Collaboration:2011ym} with energy between 360-2160 GeV. For the calculations
we use the angular dependence for $A_{eff}$ of \cite{Abbasi:2010rd}. 
\textit{Top left}:Just atmospheric background, 
152784 $\nu_{\mu} + \overline{\nu}_{\mu}$ events. \textit{Top right}:atmospheric 
background (152784 $\nu_{\mu} + \overline{\nu}_{\mu}$ events) and DM 
annihilation contribution (424 $\nu_{\mu} + \overline{\nu}_{\mu}$ events) 
from prolate Einasto profile with homogeneous annihilation cross-section 
enhancement. \textit{Middle}:
atmospheric background (152784 $\nu_{\mu} + \overline{\nu}_{\mu}$ events) and DM
annihilation contribution (332 $\nu_{\mu} + \overline{\nu}_{\mu}$ events)
from prolate Einasto profile with $\propto r^{1/4}$ annihilation 
cross-section enhancement. Numbers refer to the entire sky. We mask out the 
$\mid b \mid < 5^{\circ}$ to account for neutrinos from point sources concentrated 
on the disk and from galactic diffuse concentrated also on the disk (see text 
for more details). \textit{Bottom left}: inner $60^{\circ} \times 60^{\circ}$ for 
the atmospheric and the XDM prolate profile case. \textit{Bottom right}: same 
region as in bottom left for the case of atmospheric background and XDM 
prolate Einasto with radial dependence. The neutrinos from the GC are minimally 
enhanced.}
\label{fig:IceCubeMaps10yr}
\end{figure*}

In Fig.~\ref{fig:IceCubeMaps10yr} we give in galactic coordinates 10 yr 
mock maps for $\nu_{\mu} + \overline{\nu}_{\mu}$
upward events with energy between 360 and 2160 GeV. The energy range has been
chosen to optimize the detection of a DM signal for the specific 2.5 TeV
XDM case.

\begin{figure*}[t]
\vspace{-0.3cm}
\hspace{-0.8cm}
\includegraphics[width=3.60in,angle=0]{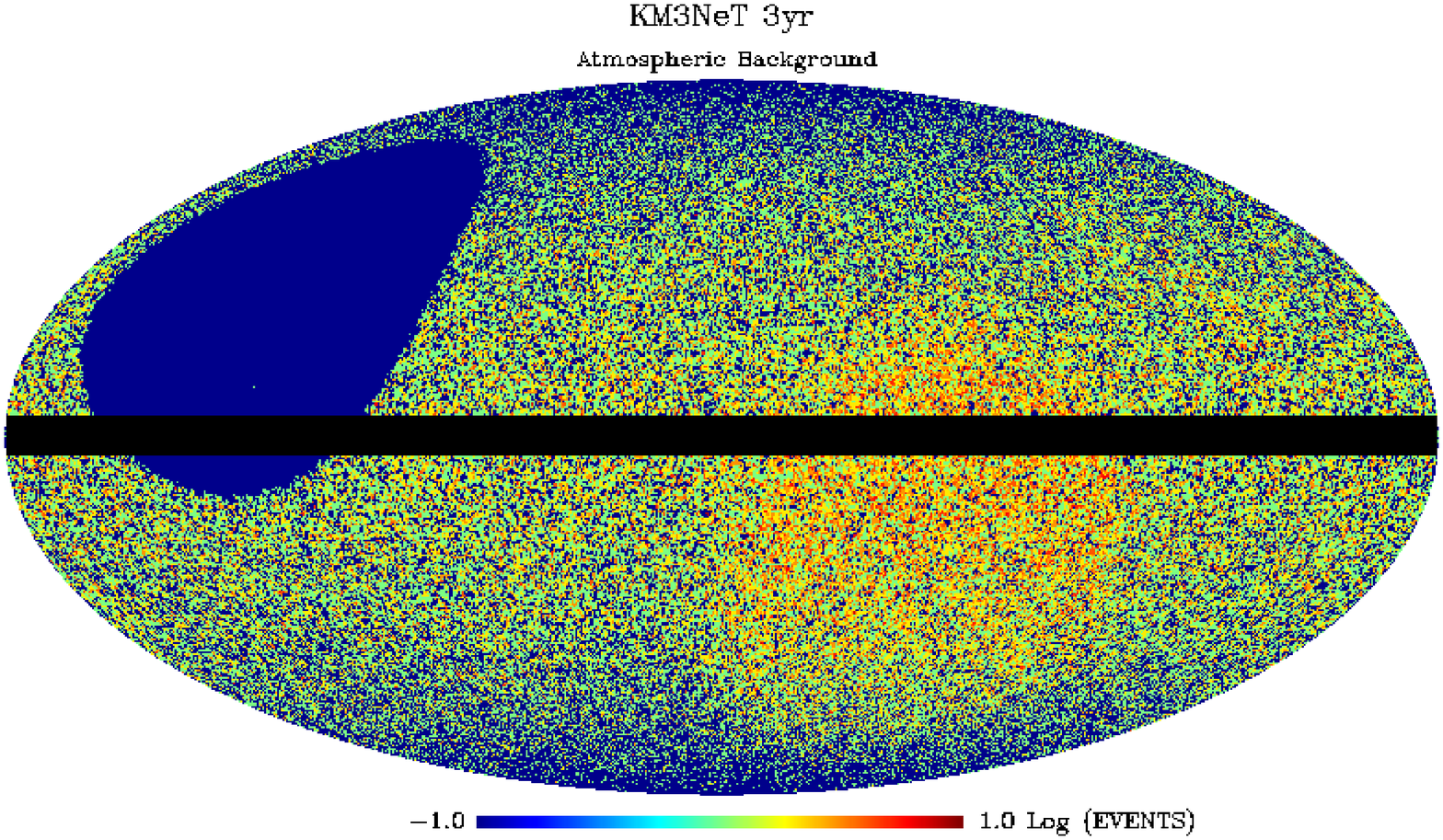}
\hspace{-0.2cm}
\includegraphics[width=3.60in,angle=0]{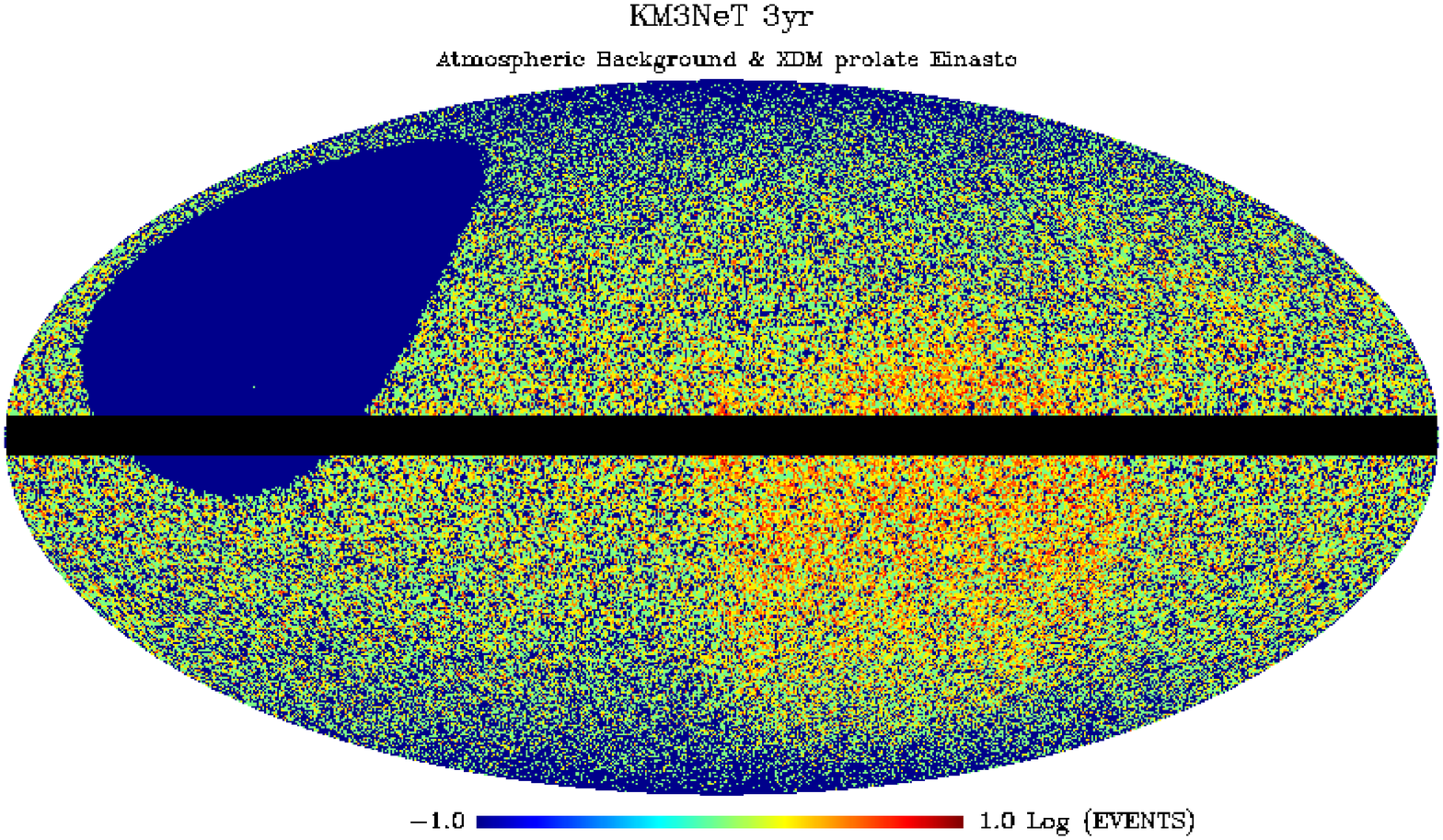} \\
\vspace{-0.3cm}
\includegraphics[width=3.60in,angle=0]{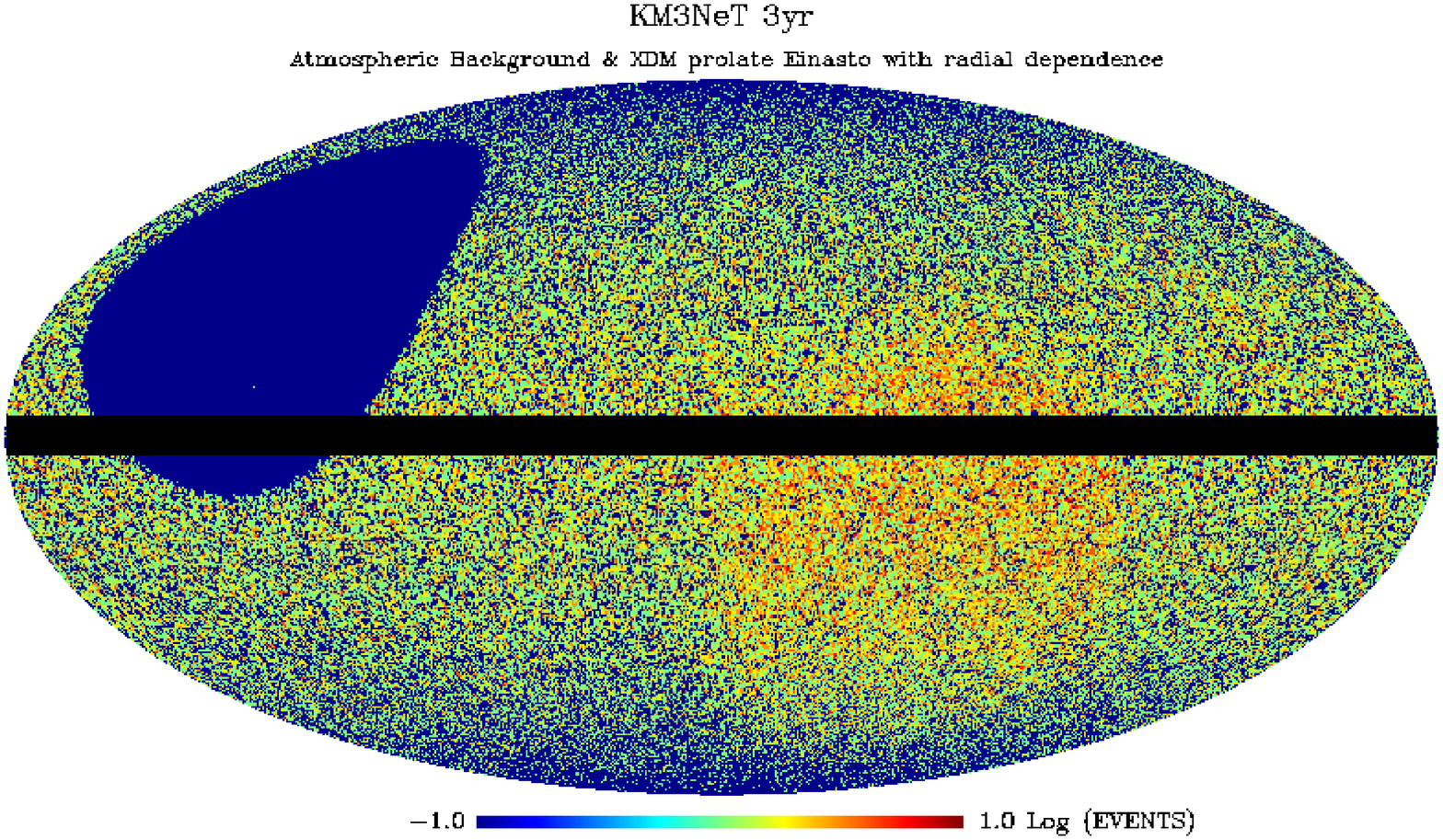}\\
\hspace{1.7cm}
\includegraphics[width=3.30in,angle=0]{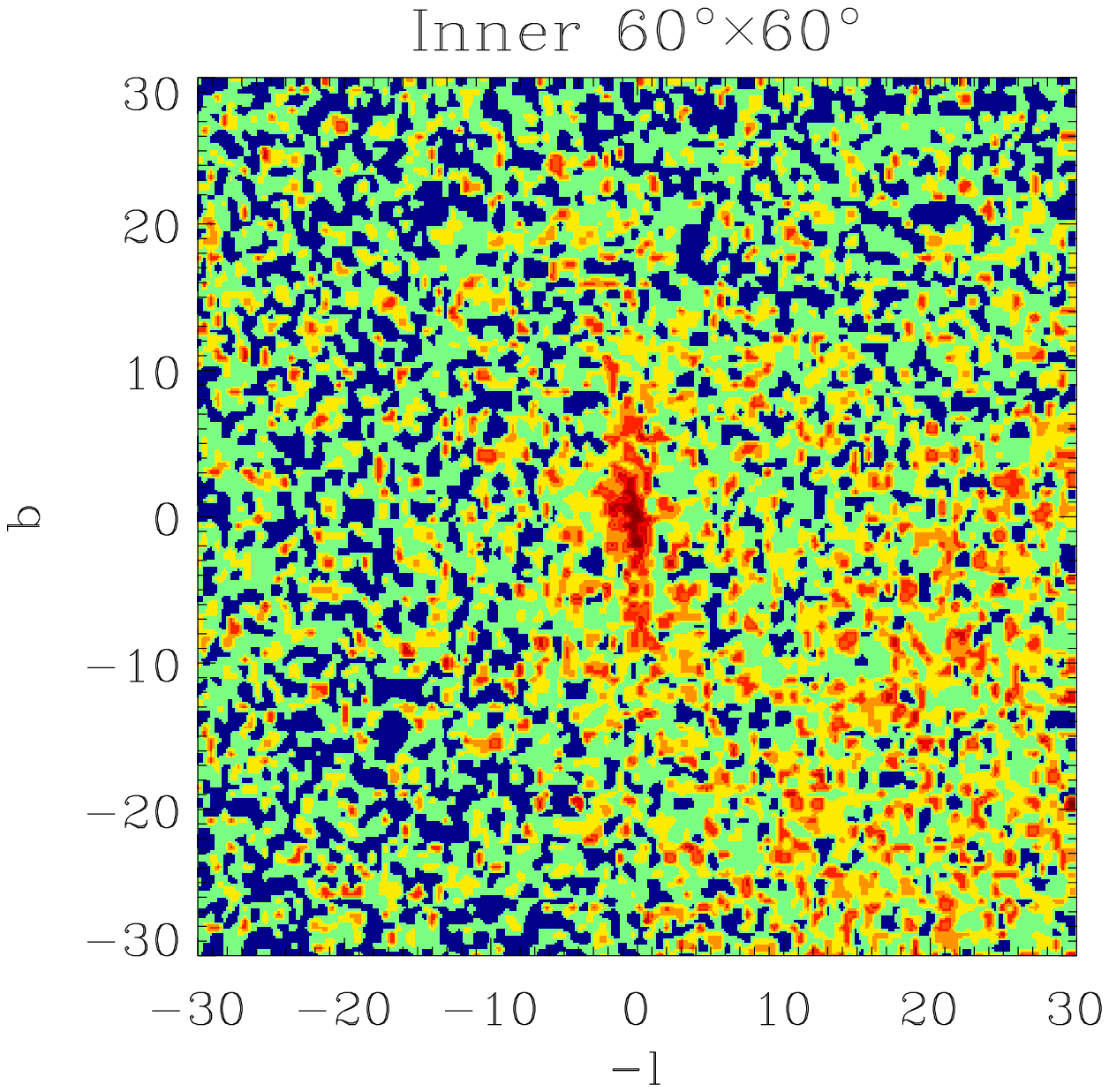}
\hspace{-1.0cm}
\includegraphics[width=3.30in,angle=0]{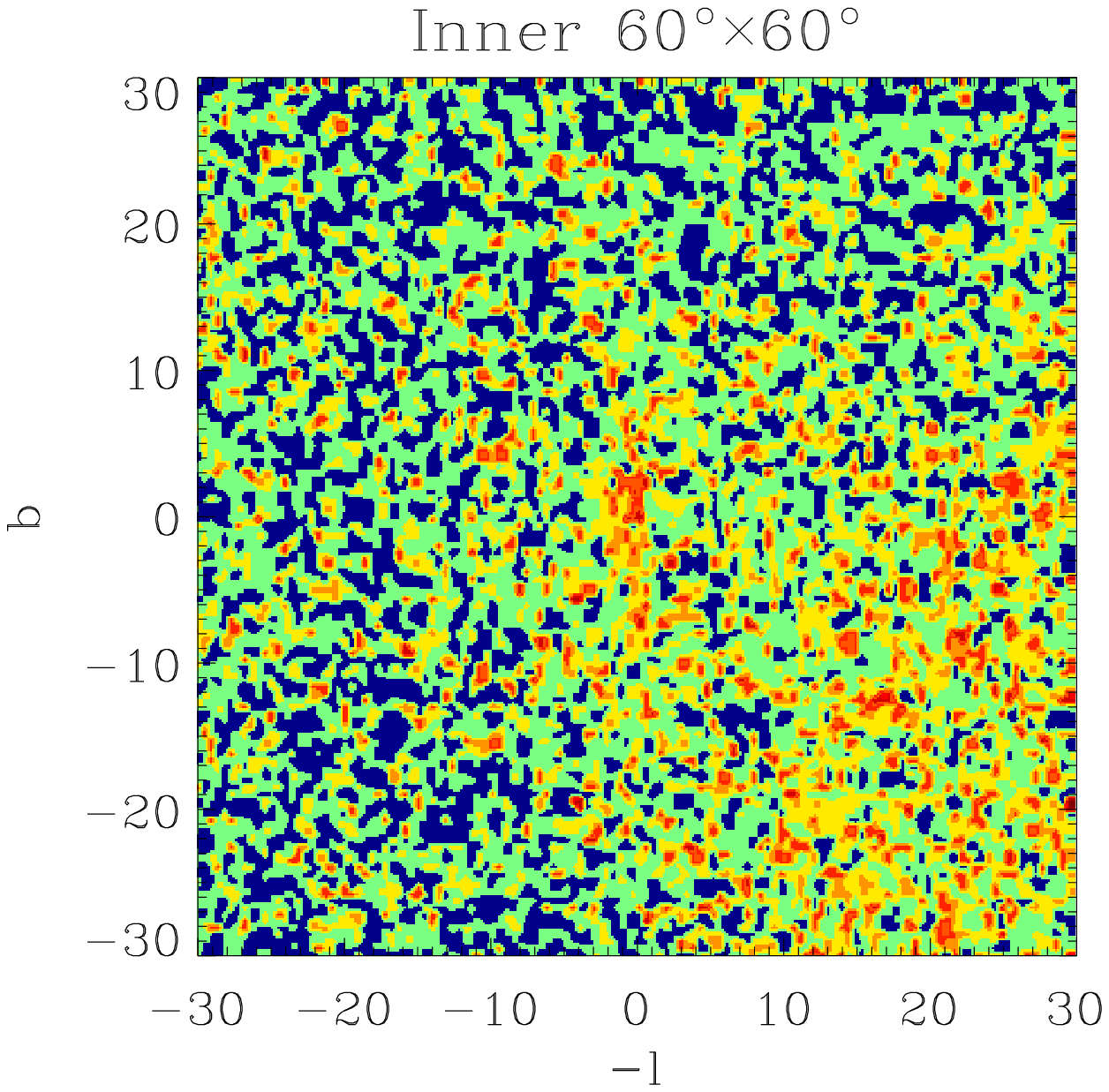}
\caption{KM3NeT $\nu_{\mu} + \overline{\nu}_{\mu}$ events
 after 3 years of data reconstructed by the HOURS \cite{Tsirigotis:2012nr} 
package with energy in the range of 360-2160 GeV. We use for KM3NeT the same
angular dependence of $A_{eff}$ as for ANTARES \cite{Heijboer:2011zz} due to their expected
similar geographic latitude location. \textit{Top left}:Just atmospheric background,
212446 $\nu_{\mu} + \overline{\nu}_{\mu}$ events. \textit{Top right}:atmospheric
background (212446 $\nu_{\mu} + \overline{\nu}_{\mu}$ events) and DM
annihilation contribution (2412 $\nu_{\mu} + \overline{\nu}_{\mu}$ events)
from prolate Einasto profile with homogeneous annihilation cross-section
enhancement. \textit{Middle}:
atmospheric background (212446 $\nu_{\mu} + \overline{\nu}_{\mu}$ events) and DM
annihilation contribution (1579 $\nu_{\mu} + \overline{\nu}_{\mu}$ events)
from prolate Einasto profile with $\propto r^{1/4}$ annihilation
cross-section enhancement. As in Fig.~\ref{fig:IceCubeMaps10yr} 
events numbers refer to the entire sky and we use the same mask of 
$\mid b \mid < 5^{\circ}$. \textit{Bottom left} and \textit{bottom right}: 
as the bottom plots of Fig.~\ref{fig:IceCubeMaps10yr} for the case of atmospheric 
background and XDM prolate Einasto without (left) and with (right) radial 
dependence. Due to its high sensitivity towards the GC and its good angular 
resolution KM3NeT would observe a clear signal from these models.}
\label{fig:KM3NeTMaps3yrRecon}
\end{figure*}

The atmospheric background $\nu_{\mu} + \overline{\nu}_{\mu}$ events that are
shown in the top left of Fig.~\ref{fig:IceCubeMaps10yr} are 152784; while the
DM events for the prolate Einasto are 424 and 332 for the case of
$\langle \sigma v \rangle \propto r^{1/4}$ for the entire sky. Those numbers of
DM events are smaller than the 1 $\sigma$ deviation (for the entire sky).
Since the morphology of the DM
signal is much different than that of the atmospheric background,  one can expect
to see a signal increase of events towards the GC (see the bottom row of
Fig.~\ref{fig:IceCubeMaps10yr}) where we show the inner $60^{\circ} \times 60^{\circ}$.
Including the $\mid b \mid < 5^{\circ}$ mask for the TeV sources (point and
diffuse) for the galactic center and disk most of this dim DM contribution is hidden.
In fact since for IceCube the neutrino events are minimally enhanced in the GC
and TeV neutrino point sources and the diffuse galactic neutrinos are also expected
to contribute in that region of the sky, a claim for DM can not be made.
Thus for the IceCube DeepCore the sensitivity towards the GC is still too low
and the angular resolution (not accounted for here) too large
($\gsim 5^{\circ}$) to provide a robust signal for that DM model.

\begin{figure*}[t]
\includegraphics[width=3.20in,angle=0]{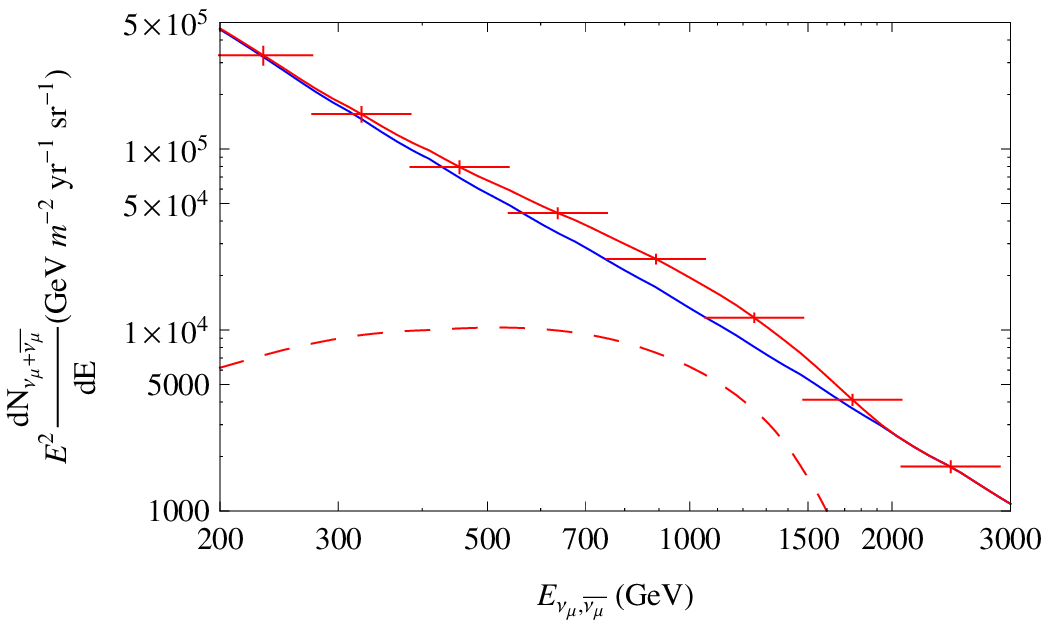}
\hspace{0.1cm}
\includegraphics[width=3.20in,angle=0]{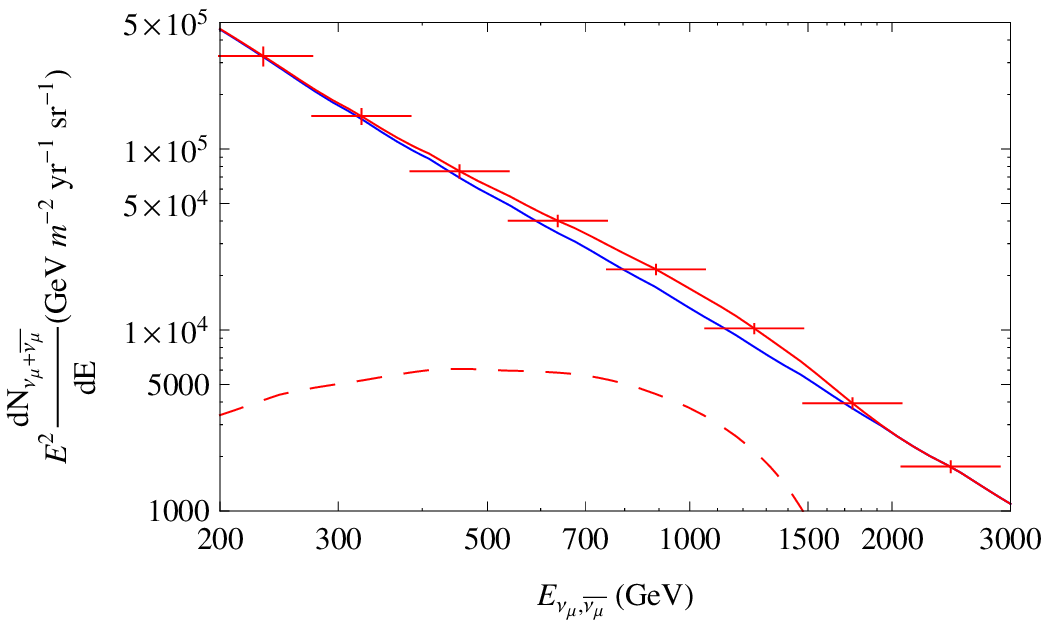}
\caption{KM3NeT $\nu_{\mu} + \overline{\nu}_{\mu}$ reconstructed spectra 
after 3 years within $5^{\circ} < \mid b \mid < 15^{\circ}$ and 
$\mid l \mid <5^{\circ}$. \textit{Blue solid line}: atmospheric
background flux, \textit{red bashed line}: DM only flux, 
\textit{red solid line}: atmospheric+DM flux. \textit{Left}: Einasto prolate 
profile with homogeneous annihilation cross-section enhancement. 
\textit{Right}: 
Einasto prolate profile with $\propto r^{1/4}$ 
with annihilation cross-section enhancement. }
\label{fig:KM3NeT3yrReconSpect}
\end{figure*}

For a km$^3$ telescope in the North Hemisphere the situation can be very different.
In Fig.~\ref{fig:KM3NeTMaps3yrRecon} we give the expected 3 yr mock maps of
reconstructed events using the HOURS simulation \cite{Tsirigotis:2012nr} for
KM3NeT\footnote{Similar results for DM and atmospheric neutrinos in KM3NeT come
from simulations as that of \cite{Lenis:2012uw}.}. In that case by comparing
Fig.~\ref{fig:KM3NeTMaps3yrRecon} top
left where we show only the atmospheric background simulated events with
Fig.~\ref{fig:KM3NeTMaps3yrRecon} top right (for combined DM Einasto prolate
and the atmospheric background), and middle (for the prolate Einasto with
$\langle \sigma v \rangle \propto r^{1/4}$ plus atmospheric), one can see a clear
indication of a signal from the 2.5 TeV XDM to muons model used to explain the
combination of \textit{Fermi} and \textit{WMAP} haze signals.
Concentrating in the inner $60^{\circ} \times 60^{\circ}$ window for the case of
simple Einasto profile the signal will be very clear even in the total counts.
Known TeV $\gamma$-ray sources such as the H.E.S.S. J1745-290
\cite{Aharonian:2004wa, vanEldik:2007yi, Aharonian:2009zk}
are also expectred to contribute in that region. Yet the angular
resolution of KM3NeT at $\sim 1$TeV is expected to be below $1^{\circ}$
\cite{Tsirigotis:2012nr}, thus given the already precise known location
of these sources (from $\gamma$-rays) it will be easy to account for
their contribution, as has been done with $\gamma$-rays H.E.S.S. data
\cite{Aharonian:2006wh} and with \textit{Fermi} data \cite{Hooper:2010mq}.

With 10 years of data with KM3NeT
the excess towards the GC will be very evident. About $8 \times 10^{3}$ DM events 
for the case of an
Einasto prolate profile and $5 \times 10^{3}$ for the Einasto prolate profile with
$\langle \sigma v \rangle \propto r^{1/4}$ will be detected.

In Fig.~\ref{fig:KM3NeT3yrReconSpect} we show the  $\nu_{\mu} + \overline{\nu}_{\mu}$
reconstructed spectra from atmospheric and from XDM to muons after 3 yrs of
observations in KM3NeT, from a window of $5^{\circ} < \mid b \mid < 15^{\circ}$
and $\mid l \mid < 5^{\circ}$, which was chosen to be optimal for a search
of a signal from DM annihilation.

For an alternative comparison to those of Fig.~\ref{fig:KM3NeTMaps3yrRecon} 
and~\ref{fig:KM3NeT3yrReconSpect} we also give in Table~\ref{tab:Sensitivities} the 
time-scale in years for either IceCube or a telescope such as KM3NeT to observe 
100 events from DM annihilation from the selected window of 
$5^{\circ} < \mid b \mid < 15^{\circ}$, $\mid l \mid < 5^{\circ}$ and energy 
range of 1.0 - 1.3 TeV around which the neutrino spectrum of DM origin peaks.
It is clear that a signal from XDM $\mu^{\pm}$ can not be probed by IceCube in any 
reasonable time-scale. Yet, a counter-part experiment in the northern hemisphere, 
with characteristics as those of the proposed KM3NeT, can detect such a signal 
within a few years. 

\begin{table}[t]
\begin{tabular}{|c||c||c|}
\hline
Signal of Interest & IceCub (yr) & KM3NeT (yr) \\
\hline \hline
XDM $\mu^{\pm}$ 2.5 TeV & 158 (294) & 3.4 (5.8) \\
\hline
 $\chi \chi \longrightarrow$ $\mu^{+}\mu^{-}$ 1.5 TeV & 38 (168) & 0.76 (3.3) \\
\hline
 $\chi \chi \longrightarrow$ $W^{+}W^{-}$ 2.0 TeV & 44 (193) & 0.86 (3.8) \\
\hline
\textit{Fermi} Bubbles & 9.0 (58) & 0.11 (0.54) \\
\hline 
\end{tabular}
\caption{The time-scale to observe 100 $\nu_{\mu}+\bar{\nu}_{\mu}$ upward events associated to 
signals of interest in IceCube (with online filter) and in KM3NeT (assuming the HOURS simulation).
In all the DM cases (XDM $\mu^{\pm}$,  $\chi \chi \longrightarrow$ $\mu^{+}\mu^{-}$ and 
$\chi \chi \longrightarrow$ $W^{+}W^{-}$), we use the same sky region of interest: 
$5^{\circ} < \mid b \mid < 15^{\circ}$, $\mid l \mid < 5^{\circ}$ (see main text for motivation)
and an energy range of 1.0-1.3 TeV,  around which these neutrino (observed) spectra peak.
For the XDM $\mu^{\pm}$ case we show results for an Einasto prolate DM profile without (with) $r^{1/4}$
velocity induced suppression to the  DM annihilation cross-section. The same Einasto prolate profile for a 
homogeneous DM annihilation cross-section is used for $\chi \chi \longrightarrow$ $\mu^{+}\mu^{-}$ and 
$\chi \chi \longrightarrow$ $W^{+}W^{-}$ with a spherical Einasto profile assumed in the results shown in the 
parentheses.
For the \textit{Fermi} Bubbles we use the entire region of the Bubbles and an energy range of 
100TeV to 1.0 PeV (100-130 TeV in parentheses).}
\label{tab:Sensitivities}
\end{table}

The uncertainty in the atmospheric neutrino flux due to uncertainties in 
the p-p collisions production cross-sections through the decay of $\pi$, 
$K$ and $\sigma$ mesons is expected to be up to $\sim 30 \%$ at TeV energies
\cite{Honda:2006qj}.
Yet, this uncertainty can not explain a change in the power law that would 
appear \textit{direction dependent} in the galactic sky. For the case in 
Fig.~\ref{fig:KM3NeT3yrReconSpect} right 
($\langle \sigma v \rangle \propto r^{1/4}$) the DM signal is too small
to be detected. 
For the case of the Einasto prolate of Fig.~\ref{fig:KM3NeT3yrReconSpect} 
left, the break at $\sim 2$ TeV is not going to be very strong, even 
in that window. Yet, a gradual hardening of the total events spectra decreasing 
from high $\mid b \mid$ towards  the galactic disk (excluding the
$\mid b \mid < 5^{\circ}$ region), will be an indication of a signal from DM 
annihilation in the main DM halo. We expect that the power law of total 
reconstructed $\nu_{\mu}$, $\overline{\nu}_{\mu}$ events with energy between 
300 GeV and 1.5 TeV \textit{will become harder by 0.3} from the window of 
$\mid b \mid > 50^{\circ}$, $\mid l \mid < 5^{\circ}$  to that of 
$5^{\circ} < \mid b \mid < 15^{\circ}$, $\mid l \mid < 5^{\circ}$.

\section{Dark Matter annihilation towards the GC, other channels}
\label{sec:ODMA}

Apart from the question on explaining the \textit{Fermi} and \textit{WMAP} 
haze signals via DM annihilation, the general connection of DM searches in 
neutrinos has received some attention in the recent years either from 
annihilation towards the 
GC \cite{Erkoca:2010qx, Erkoca:2010vk, Heros:2010ss} or from the DM 
annihilation captured in the Sun 
\cite{IceCube:2011ae, Arguelles:2012cf, Barger:2010ng, Mandal:2009yk}.  

\begin{figure*}[t]
\includegraphics[width=3.20in,angle=0]{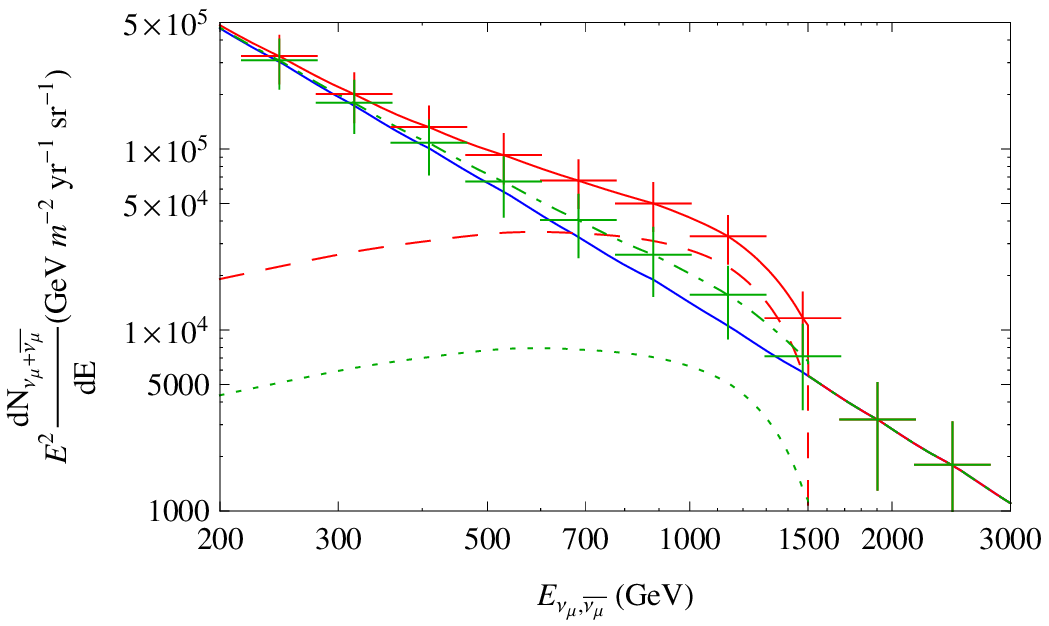}
\hspace{0.1cm}
\includegraphics[width=3.20in,angle=0]{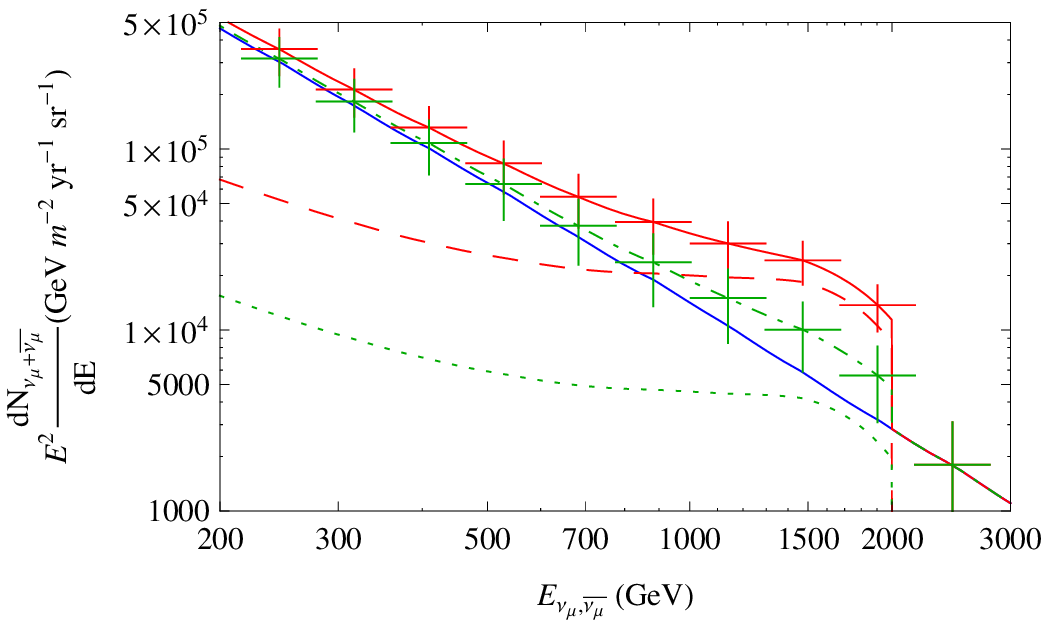}
\caption{IceCube DeepCore $\nu_{\mu} + \overline{\nu}_{\mu}$ reconstructed spectra 
after 3 years in the window of $5^{\circ} <\mid b \mid <15^{\circ}$ and
$\mid l \mid <5^{\circ}$. \textit{Blue solid line}: atmospheric
background flux, \textit{red dashed} and \textit{green dotted lines}: DM only flux,
\textit{red solid} and \textit{green dashed-dotted lines}: atmospheric+DM flux. 
\textit{Left}:
$\chi \chi \longrightarrow \mu^{+}\mu^{-}$ Einasto prolate profile
(\textit{red dashed/solid lines})
, Einasto spherical profile (\textit{green dotted/dashed-dotted lines}) with homogeneous
annihilation
cross-section enhancement. \textit{Right}: $\chi \chi \longrightarrow W^{+}W^{-}$ 
Einasto prolate (\textit{red dashed/solid lines})
, Einasto spherical (\textit{green dotted/dashed-dotted lines}) with homogeneous annihilation
cross-section enhancement.}
\label{fig:ICDC3yrReconSpectMuonsWW}
\end{figure*}
 
\begin{figure*}[t]
\includegraphics[width=3.20in,angle=0]{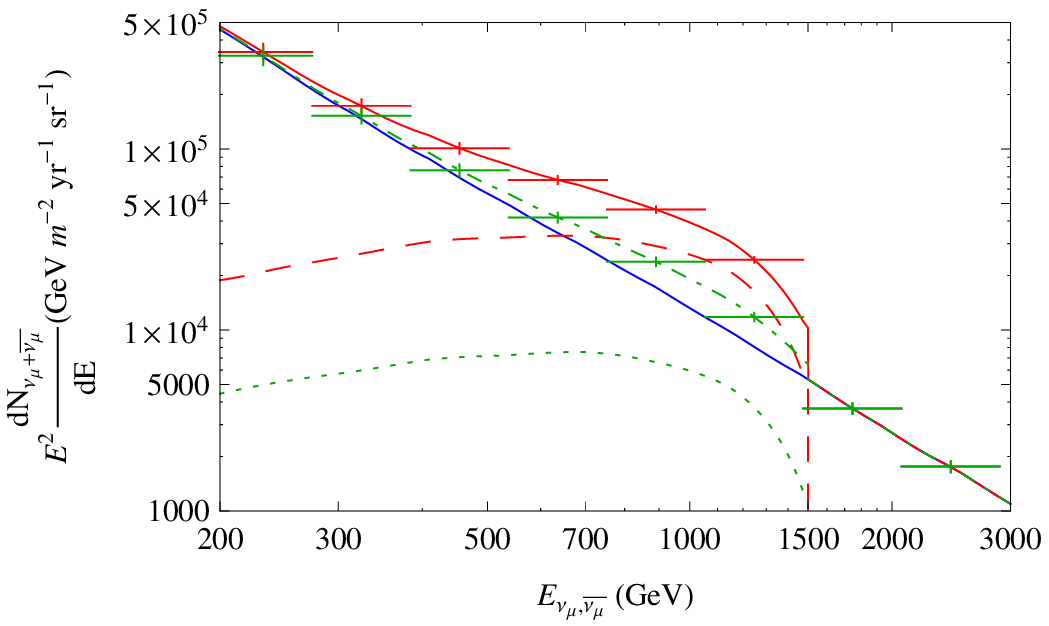}
\hspace{0.1cm}
\includegraphics[width=3.20in,angle=0]{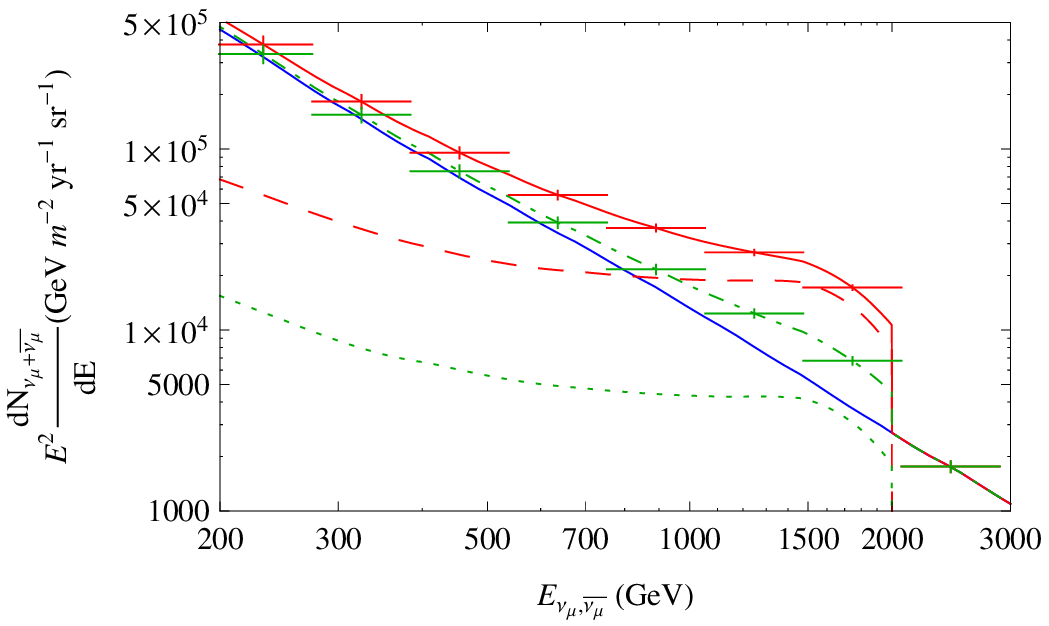}
\caption{KM3NeT $\nu_{\mu} + \bar{\nu_{\mu}}$ reconstructed spectra 
after 3 years in the window of $5^{\circ} < \mid b \mid < 15^{\circ}$ and
$\mid l \mid <5^{\circ}$. \textit{Blue solid line}: atmospheric
background flux, \textit{red bashed} and \textit{green dotted lines}: DM only flux,
\textit{red solid} and \textit{green dashed-dotted lines}: atmospheric$+$DM flux. 
\textit{Left}: 
$\chi \chi \longrightarrow \mu^{+}\mu^{-}$ Einasto prolate profile 
(\textit{red dashed/solid lines})
, Einasto spherical profile (\textit{green dotted/dashed-dotted lines}) with homogeneous 
annihilation
cross-section enhancement. \textit{Right}: $\chi \chi \longrightarrow W^{+}W^{-}$ 
Einasto prolate (\textit{red dashed/solid lines})
, Einasto spherical (\textit{green dotted/dashed-dotted lines}) with homogeneous annihilation
cross-section enhancement.}
\label{fig:KM3NeT3yrReconSpectMuonsWW}
\end{figure*}
The \textit{Fermi} and HESS $e^{-}+e^{+}$ CR flux
\cite{Abdo:2009zk, Collaboration:2008aaa, Aharonian:2009ah} and
\textit{PAMELA} positron \cite{Boezio:2008mp} excesses apart from the
XDM models discussed in~\ref{sec:LDMA} can be explained by a variety
of other phenomenological channels
\cite{Cirelli:2008jk, Bergstrom:2008gr, Cholis:2008hb, Cholis:2008wq, Zhang:2008tb}.

For standard phenomenological models
$\chi \chi \longrightarrow W^{+}W^{-}$, $\chi \chi \longrightarrow b \bar{b}$
strong limits have been placed using SUPER-K and IceCude observations
towards the Sun (see for example some recent works of
\cite{Arguelles:2012cf, Barger:2010ng, Mandal:2009yk}).

Among the many models we choose to show results for the phenomenological
models of DM annihilating directly to muons
($\chi \chi \longrightarrow \mu^{+}\mu^{-}$) or directly to W bosons
($\chi \chi \longrightarrow W^{+}W^{-}$), where electroweak corrections
-especially important for the former channel- have been included
\cite{Ciafaloni:2010ti, Cirelli:2010xx}.
These two channels are distinct between each other since for the one
annihilating to muons high energy neutrinos come from the decay of the
boosted muons, while for the $\chi \chi \longrightarrow W^{+}W^{-}$
neutrinos are produced with a softer overall spectrum but at significantly
higher multiplicity.

In Fig.~\ref{fig:ICDC3yrReconSpectMuonsWW} we show for the window of
$5^{\circ} < \mid b \mid < 15^{\circ}$, $\mid l \mid < 5^{\circ}$
the expected reconstructed upward fluxes in IceCube DeepCore of
$\nu_{\mu} + \overline{\nu}_{\mu}$, for $\chi \chi \longrightarrow \mu^{+}\mu^{-}$
with $m_{\chi} = 1.5$ TeV (left) and for $\chi \chi \longrightarrow W^{+}W^{-}$
with $m_{\chi} = 2.0$ TeV (right).

The masses and annihilation cross sections are chosen to fit the
\textit{Fermi}, HESS and \textit{PAMELA} leptonic excesses given in
\cite{Cholis:2008wq}. We show results for two cases of DM profiles,
the Einasto prolate profile of eq.~\ref{eq:ProlateEin} (red lines) and
the spherical Einasto profile (green lines). As can be seen for the more
optimistic (to give a clear DM signal) prolate profile, a signal can be seen in
IceCube DeepCore within 3 yrs of data for both channels. For the less
optimistic spherical Einasto profile only in the case of
$\chi \chi \longrightarrow W^{+}W^{-}$ a weak break at $\simeq$ 2 TeV
may be seen.
 
\begin{figure*}[t]
\vspace{-0.3cm}
\hspace{-0.8cm}
\includegraphics[width=3.60in,angle=0]{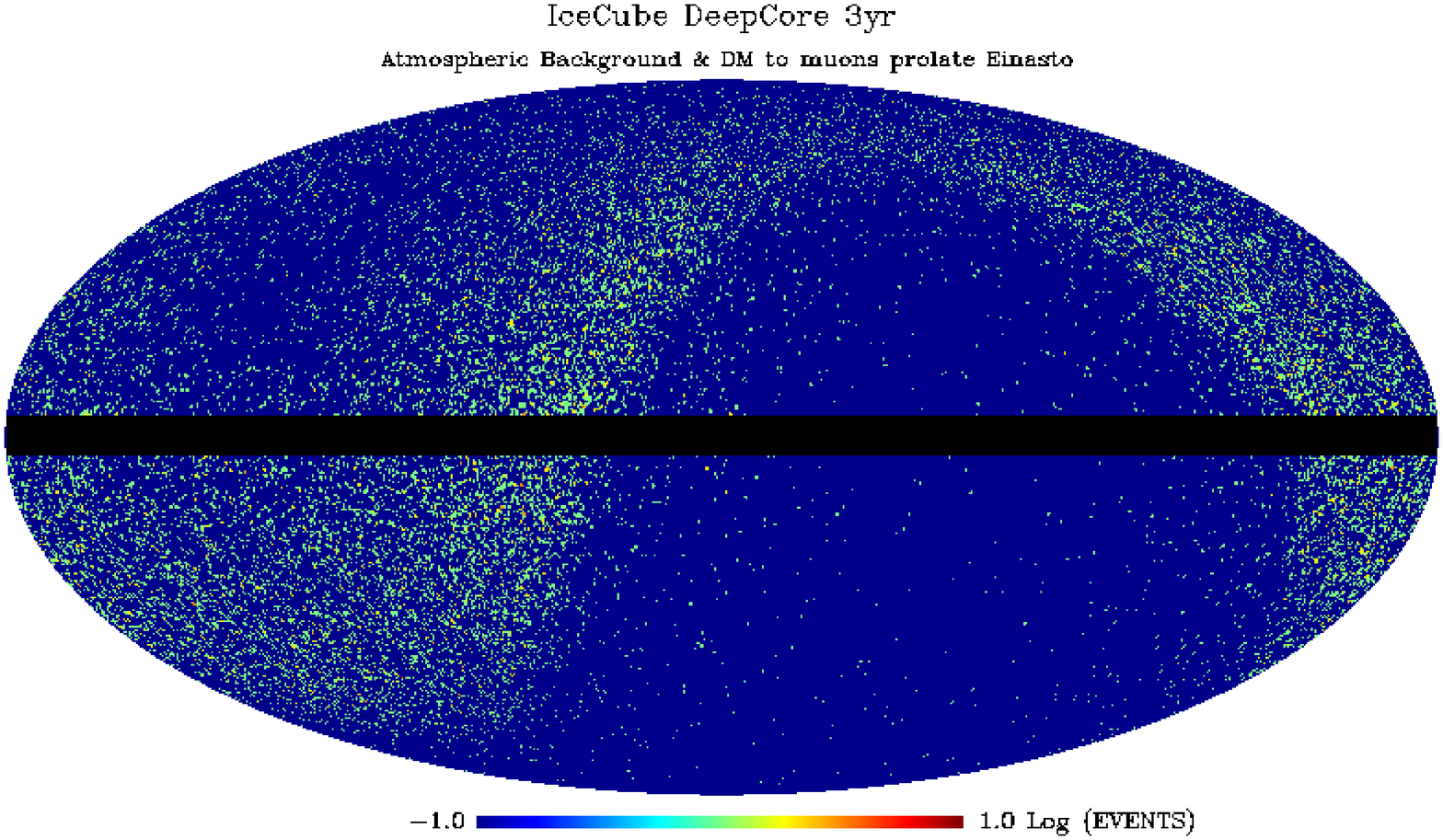}
\hspace{-0.2cm}
\includegraphics[width=3.60in,angle=0]{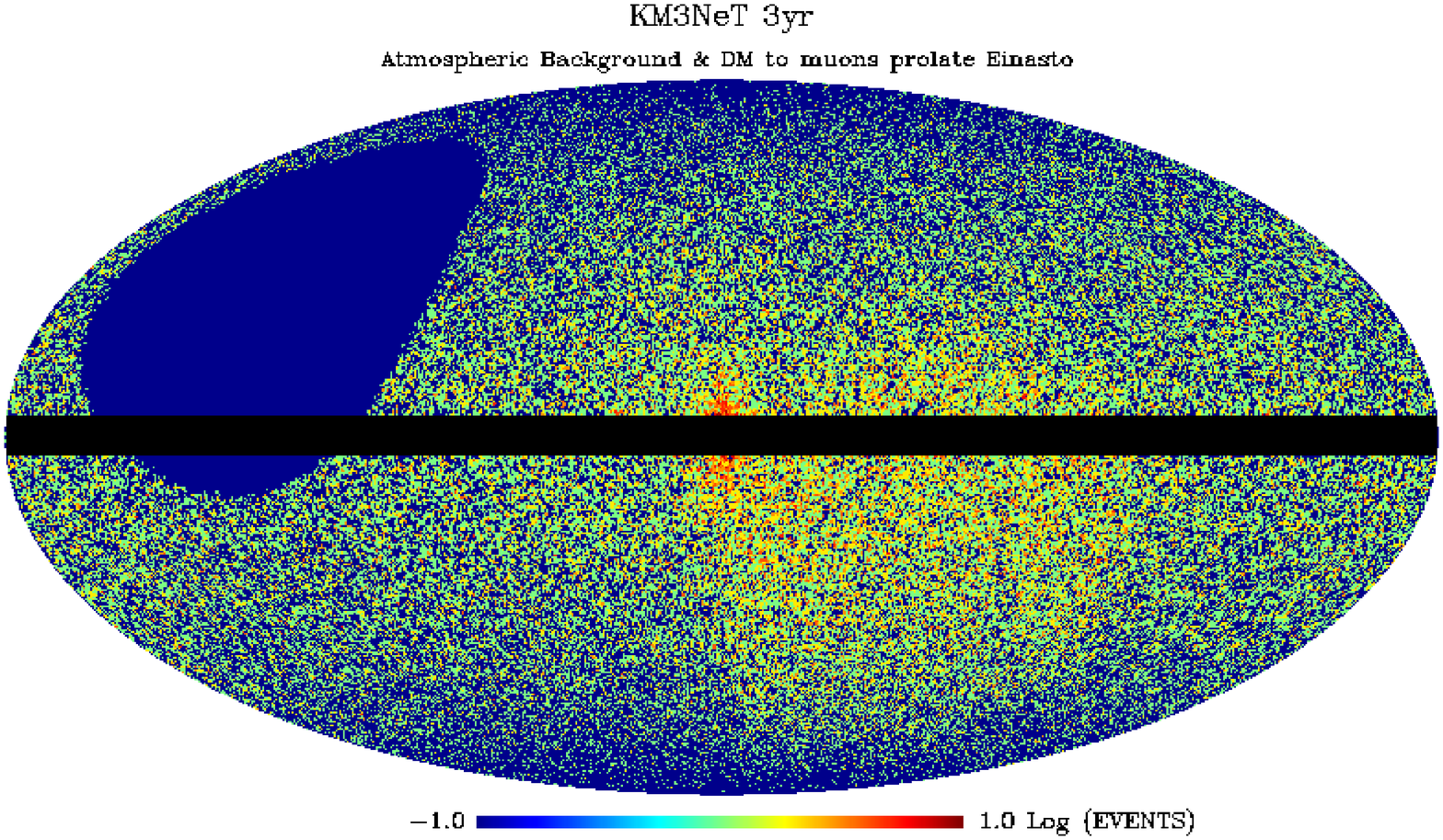} \\
\hspace{1.7cm}
\includegraphics[width=3.30in,angle=0]{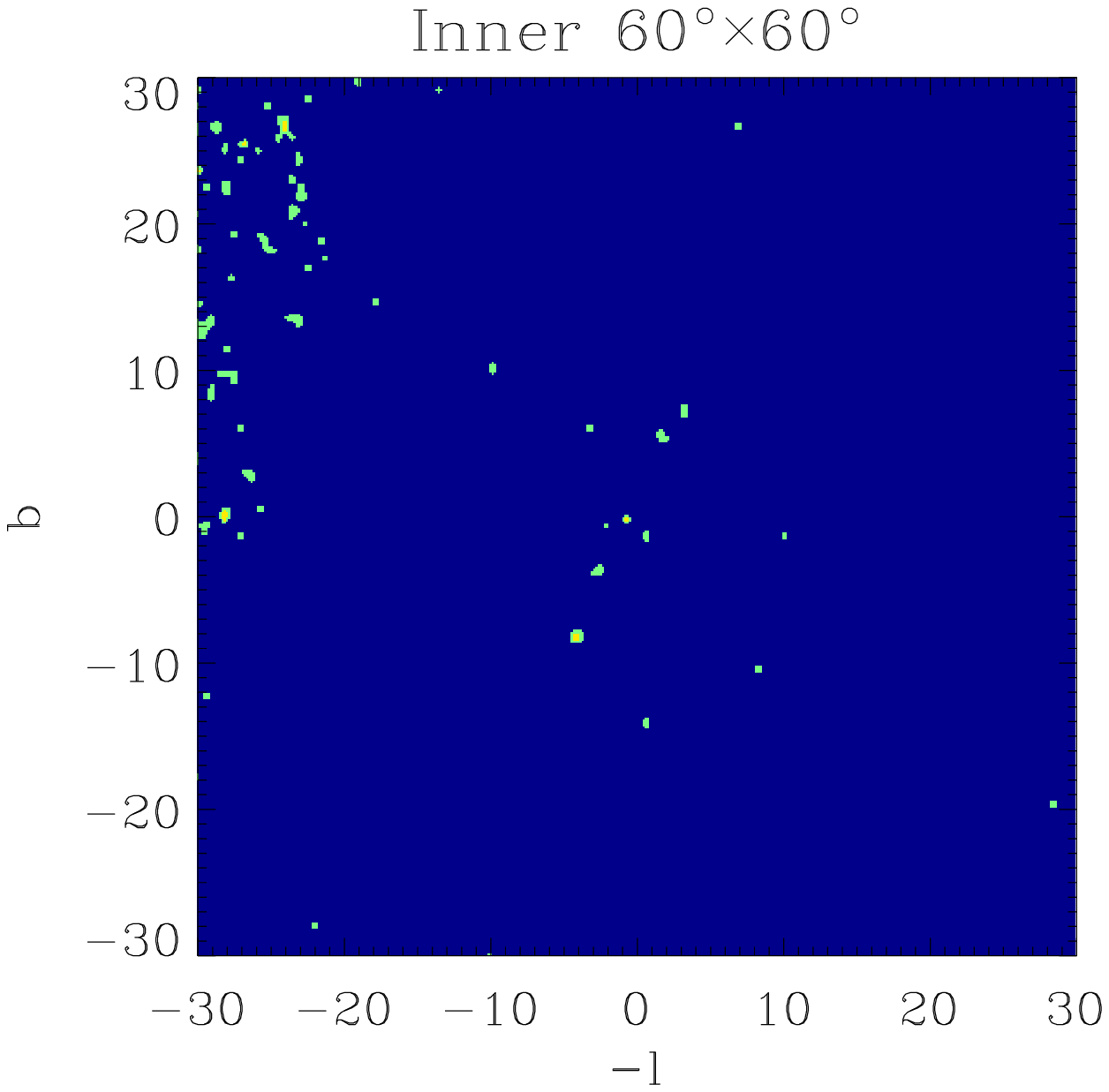}
\hspace{-1.0cm}
\includegraphics[width=3.30in,angle=0]{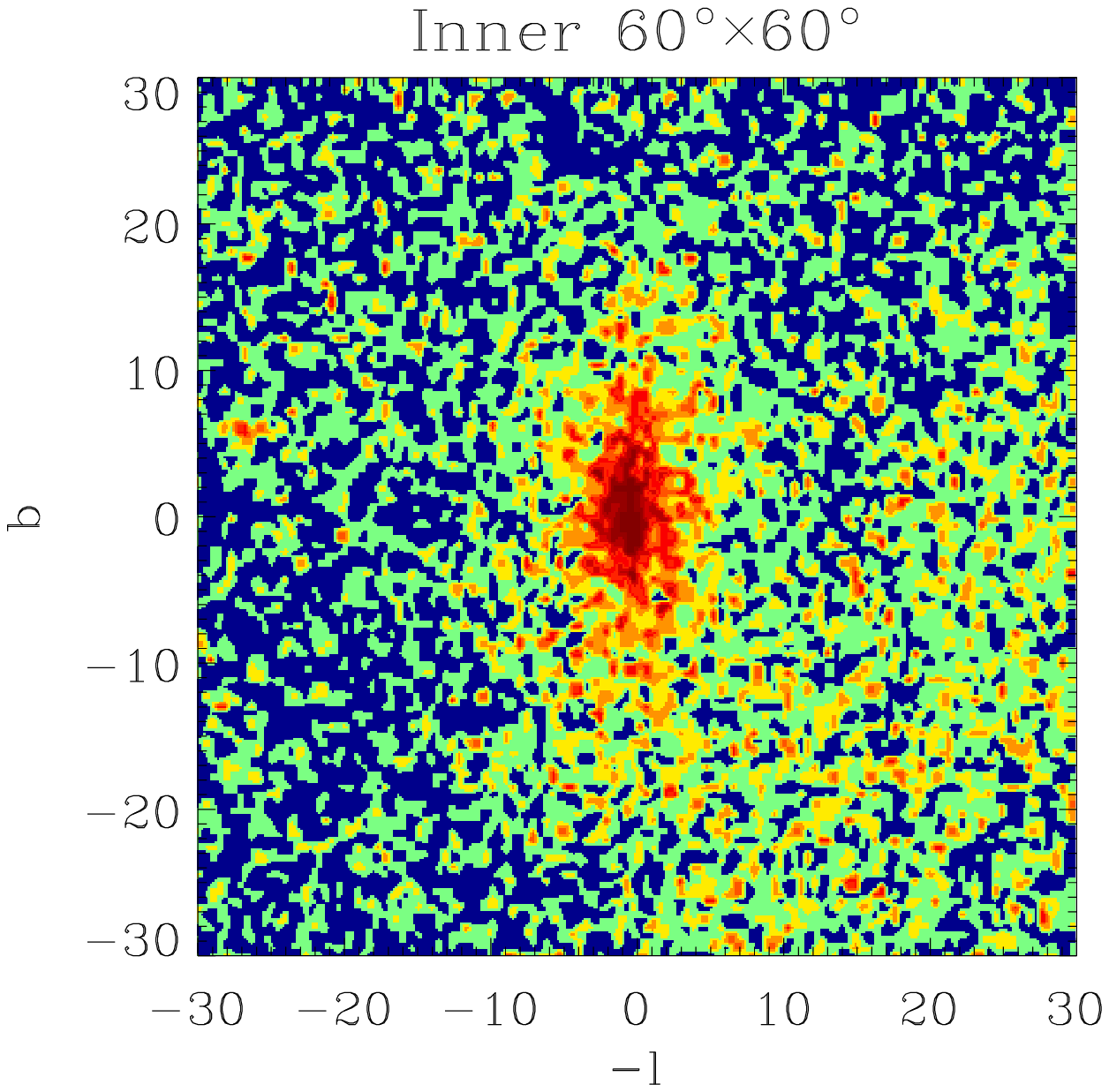}
\caption{$\nu_{\mu} + \overline{\nu}_{\mu}$ events with energy between 500 GeV and 1.5 TeV 
from DM annihilation of $M_{\chi}=1.5$ TeV
$\chi \chi \longrightarrow \mu^{+}\mu^{-}$ Einasto prolate profile. 
\textit{Top Left}: With IceCube
DeepCore in 3 yr, with ``Online Filter'' atmospheric
background (22477 $\nu_{\mu} + \overline{\nu}_{\mu}$ events) and
contribution from DM (368 $\nu_{\mu} + \overline{\nu}_{\mu}$ events).
\textit{Top Right}: With KM3NeT in 3yr using HOURS reconstruction technique,
atmospheric background (138560 $\nu_{\mu} + \overline{\nu}_{\mu}$ events) and
contribution from DM (6482 $\nu_{\mu} + \overline{\nu}_{\mu}$ events). 
As in Fig.~\ref{fig:IceCubeMaps10yr}
events numbers refer to the entire sky and we use the same mask of
$\mid b \mid < 5^{\circ}$. \textit{Bottom left} and \textit{bottom right}:
Zooming in the $60^{\circ} \times 60^{\circ}$ window for the 
IceCube (left) and KM3NeT (right) maps of the top row. Even in IceCube
some excess of events is expected to be seen towards the GC. With KM3NeT 
sensitivity and angular resolution a clear signal from that model will
be observed or strong constraints will be placed.}
\label{fig:DMmuonsDetect}
\end{figure*}
Yet, in KM3NeT -that is near to optimal for searching for such a signal-
after 3 yr of data the reconstructed spectra in the same window will be
measured with much greater statistics as is shown in
Fig.~\ref{fig:KM3NeT3yrReconSpectMuonsWW}. Apart from the case of
$\chi \chi \longrightarrow \mu^{+}\mu^{-}$ with a spherical Einasto profile,
all other cases give a clear break at the respective mass of the DM particle.
For the case of $\chi \chi \longrightarrow \mu^{+}\mu^{-}$ in the spherical
Einasto an indication of a signal will be the smooth hardening of the spectrum
as one moves (using the same longitude range) from the high $\mid b \mid$
towards the disk as discussed in section~\ref{sec:LDMA}. In the specific case
at $\mid b \mid > 40^{\circ}$ the power law is going to be $\simeq 4.0$ while
at the $5^{\circ} < \mid b \mid < 15^{\circ}$ it is going to be $\simeq 3.2$
between energies of 300 GeV and 1.0 TeV. For a more cored profile the 
difference in the power law between the $\mid b \mid > 40^{\circ}$ region 
and the $5^{\circ} < \mid b \mid < 15^{\circ}$ region of the sky, is going to 
be only by 0.2.

An alternative indication of how much better a km$^3$ telescope in the
North Hemisphere is going to be compared to IceCube DeepCore for that type
of searches, is given in Fig.~\ref{fig:DMmuonsDetect}. In
Fig.~\ref{fig:DMmuonsDetect} we show the reconstructed $\nu_{\mu} + \overline{\nu}_{\mu}$
within 500 GeV and 1.5 TeV for the $m_{\chi}=1.5$ TeV,
$\chi \chi \longrightarrow \mu^{+}\mu^{-}$ case. For the IceCube DeepCore
(left) and the KM3NeT (right) using the reconstructed HOURS simulation
\cite{Tsirigotis:2012nr}.
For the entire DM halo (ignoring substructure) we show 368
$\nu_{\mu} + \overline{\nu}_{\mu}$ events (3 yr) in IceCube DeepCore and
6482 for KM3NeT (3yr), with the equivalent background events to be $2.2 \times 10^{4}$
and $1.4 \times 10^{5}$. In  Table~\ref{tab:Sensitivities} we also give for both the 
$\chi \chi \longrightarrow \mu^{+}\mu^{-}$ and the $\chi \chi \longrightarrow W^{+}W^{-}$
channels the time-scale needed for each experiment to observe 100 
$\nu_{\mu} + \overline{\nu}_{\mu}$ upward events from the direction of 
$5^{\circ} < \mid b \mid < 15^{\circ}$, $\mid l \mid < 5^{\circ}$ and in the energy
range of 1.0-1.3 TeV (see also Figs.~\ref{fig:ICDC3yrReconSpectMuonsWW} 
and~\ref{fig:KM3NeT3yrReconSpectMuonsWW}). 

We note that current limits on the muon channel (strongest of which come 
from dwarf spheroidal galaxies (dSph)\cite{Ackermann:2011wa, Cholis:2012am}) 
can not exclude a cross 
section of $\langle \sigma v \rangle \simeq 9.5 \times 10^{-24}$ cm$^{3}$s$^{-1}$ 
used here for the 1.5 TeV mass.
For the case of $\chi \chi \longrightarrow W^{+}W^{-}$ the cross section was taken 
to be $\simeq 7 \times 10^{-23}$ cm$^{3}$s$^{-1}$. This cross section is a factor of 
$\simeq 5$ times higher that conservative limits coming from dSph or antiprotons 
\cite{Cirelli:2008id, Donato:2008jk, Cholis:2010xb}. For the most optimistic cross 
sections that are still allowed by $\bar{p}$s and $\gamma$-rays from dSphs, only for 
a very optimized DM halo profile (such as a prolate or in general cuspy profile), 
can there be a
 signal that would be detected. Thus, the neutrino searches for hadronic channels 
are deemed not optimal given how strong the constraints are that can be drawn from the local 
$\bar{p}$ flux and from $\gamma$-ray searches for these channels. For leptonic to 
mainly muons channels though, $\bar{p}$ and $\gamma$-s provide weak constraints and 
neutrinos can provide a useful alternative search channel.
    
\section{Hadronic Scenario: Neutrinos from the Fermi Bubbles }
\label{sec:HSFB}

As discussed in \cite{Crocker:2010dg} a possible explanation of the 
\textit{Fermi} Bubbles signal \cite{Su:2010qj} is that of copious and long 
time-scale star formation in the galactic center giving CR protons with 
energies up to the PeV scale. These CR protons are transferred up to a 
distance of 10 kpc away from the galactic disk due to strong winds 
\cite{Crocker:2010dg}. In that scenario the $\gamma$-rays composing the 
\textit{Fermi} Bubbles will come from the decay chains of boosted mesons 
produced in p-p collisions. The same process will produce neutrinos with 
energies up to the cut-off energy of these hard CR protons. 
Similarly these 
processes take place in the atmosphere producing the equivalent background.

The CR protons entering the atmosphere have a significantly softer 
spectrum ($\simeq$ 2.67 above 300 GeV) measured most recently from 
\cite{Adriani:2011cu, Yoon:2011zz}
\footnote{Older measurements suggest a value for the power law of the 
differential spectrum closer to 2.7-2.8 at the multi TeV scale 
\cite{1998ApJ...502..278A, 2005ApJ...628L..41D} 
thus slightly softer spectra}, compared to those responsible for the 
Bubbles that have a spectrum described by \cite{Crocker:2010dg}:
\begin{equation}
\frac{dN_{p}}{dE_{p}} dE = N_{0} E_{p}^{-2.1} exp\left[-E_{p}/E_{p_{0}} \right],
 \label{eq:protonBubbles}
\end{equation}
with $E_{p_{0}}$ the cut-off energy $\sim$ PeV. Thus we can expect to see the 
neutrino component from the Bubbles at high energies 
\cite{Crocker:2010qn, Vissani:2011ea, Lunardini:2011br}.

Another difference between the atmospheric neutrino background and the possible 
neutrino signal from the \textit{Fermi} Bubbles, is that the CR protons entering 
the atmosphere due to column densities of matter $\simeq$ kg$\,$cm$^{-2}$,
produce extensive showers that can reach for the most energetic protons 
up to $10^{10}$ particles at peak number \cite{Abraham:2009bc, Matthews:2005sd, Longair} 
while for the CR protons
at the Bubbles region one expects much lower column densities \footnote{As a 
comparison towards the GC the ISM column density is $\simeq 10^{-4}$ kg$\,$cm$^{-2}$}.
Thus the possible \textit{Fermi} Bubbles neutrinos can not come from extensive 
showers, where products (protons mainly) from the hadronization of the initial 
p-p collision would then hit on new target protons. 
Rather, the neutrinos will come from the decay chains of the  hadronization 
products related to a single hard p-p inelastic collision.  
This is an additional reason why the neutrinos from the Bubbles to have a 
harder spectrum than the atmospheric ones. 

For the p-p inelastic processes the neutrinos are mainly produced from charged pion 
decays. For the neutrino spectra we follow the parametrization of \cite{Kelner:2006tc},
that was based on SIBYLL \cite{Fletcher:1994bd} simulations of p-p collisions 
and is optimal at energies above 100 GeV that we care for.
 
\begin{figure}
\hspace{-0.6cm}
\includegraphics[width=3.60in,angle=0]{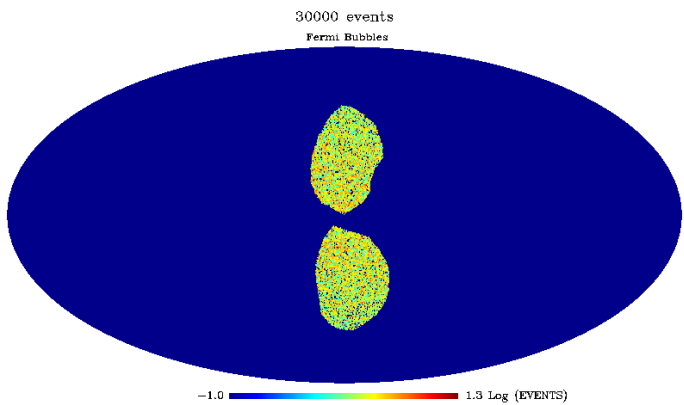}
\caption{\textit{Fermi} Bubbles, $3\times 10^{4}$ simulated events.}
\label{fig:Bubbles}
\end{figure}

The neutrinos coming from the Bubbles will have the same morphology as the 
$\gamma$-rays, which is relatively flat in longitude and latitude with clear 
edges \cite{Su:2010qj}. 
However, it may not be trivial for the CR protons and the ISM target material transferred
with them by the galactic winds, to cause such a flat morphology in $l$ and $b$, 
at $\gamma$-rays; following the assumptions of \cite{Crocker:2010dg} for the 
$\gamma$-rays we will take the morphology of the neutrinos 
shown in Fig.~\ref{fig:Bubbles} to be flat within the Bubbles region, with clear 
edges as in the \textit{Fermi} Bubbles signal of \cite{Su:2010qj}.     

As can be seen by comparing the morphology of Fig.~\ref{fig:Bubbles} to those of 
 Fig.~\ref{fig:DMmaps}, that are for neutrinos from DM scenarios which fit 
the \textit{Fermi} and \textit{WMAP} haze, the two types of morphologies 
are distinctively  different. Moreover the neutrino spectra and fluxes as we will 
show differ dramatically.
In both Figs.~\ref{fig:DMmaps} and~\ref{fig:Bubbles} we show the same number 
of neutrino events without specifying the energy range or period of observation
and we are not taking into account that any actual neutrino telescope has (or will 
have) a strong angular dependence of its sensitivity. 
This is done to ``spotlight'' the different morphologies of the possible neutrino
signals. Even after taking into account the specific properties of neutrino
telescopes that we show results for, the different morphologies lead to searching
for these signals in different parts of the sky. 

The total energy stored in the CR protons in the Bubbles is estimated 
to be $\sim 10^{56}$erg due to an estimated averaged $10^{39}$erg/s 
of injected power to hard CR protons transferred from the GC via galactic 
winds in the \textit{Fermi} Bubbles regions. This process is estimated 
to have been ongoing for a timescale of multi Gyrs \cite{Crocker:2010dg}.
These assumptions can result in a quasi-steady state injected energy 
from protons to $\gamma$, $e^{\pm}$ and $\nu$ of 
$\dot{Q}_{p} \simeq 3.6 \times 10^{38}$ erg/s from 10 GeV to 1 PeV
\cite{Crocker:2010dg}. Of that power from approximate equal partitions of 
energy to $\pi^{0}$, $\pi^{+}$ and $\pi^{-}$, 1/3 goes to $\gamma$s
giving the better estimated power in the \textit{Fermi} Bubbles
of $\simeq 2 \times 10^{37}$ erg/s for $\gamma$-ray with energy 
1-100 GeV \cite{Su:2010qj}, or after extrapolation of the  $\gamma$-ray 
spectrum, $\simeq 1.2 \times 10^{38}$ erg/s for energies between 10 GeV and 1 TeV.  
Also from the energy equipartition to the product pions we can estimate that 
the power to neutrinos is $\sim 2 \times 10^{38}$ erg/s inside the Bubbles.
We take the power to neutrinos to be $1 \times 10^{38}$ erg/s, to account
for an overestimation by a factor of 2 on the $\gamma$-ray luminosity 
of the bubbles by \cite{Crocker:2010dg}. 
Assuming isotropic emission of $\nu$s inside the bubbles and a mean distance 
squared $D^{2} \simeq R_{sun}^{2} \simeq 8.5^{2}$ kpc$^{2}$ from us, we can
estimate their flux. Their observed morphology as shown in 
Fig.~\ref{fig:Bubbles} is isotropic within the Bubbles.

\begin{figure*}[t]
\vspace{-0.3cm}
\hspace{-0.8cm}
\includegraphics[width=3.10in,angle=0]{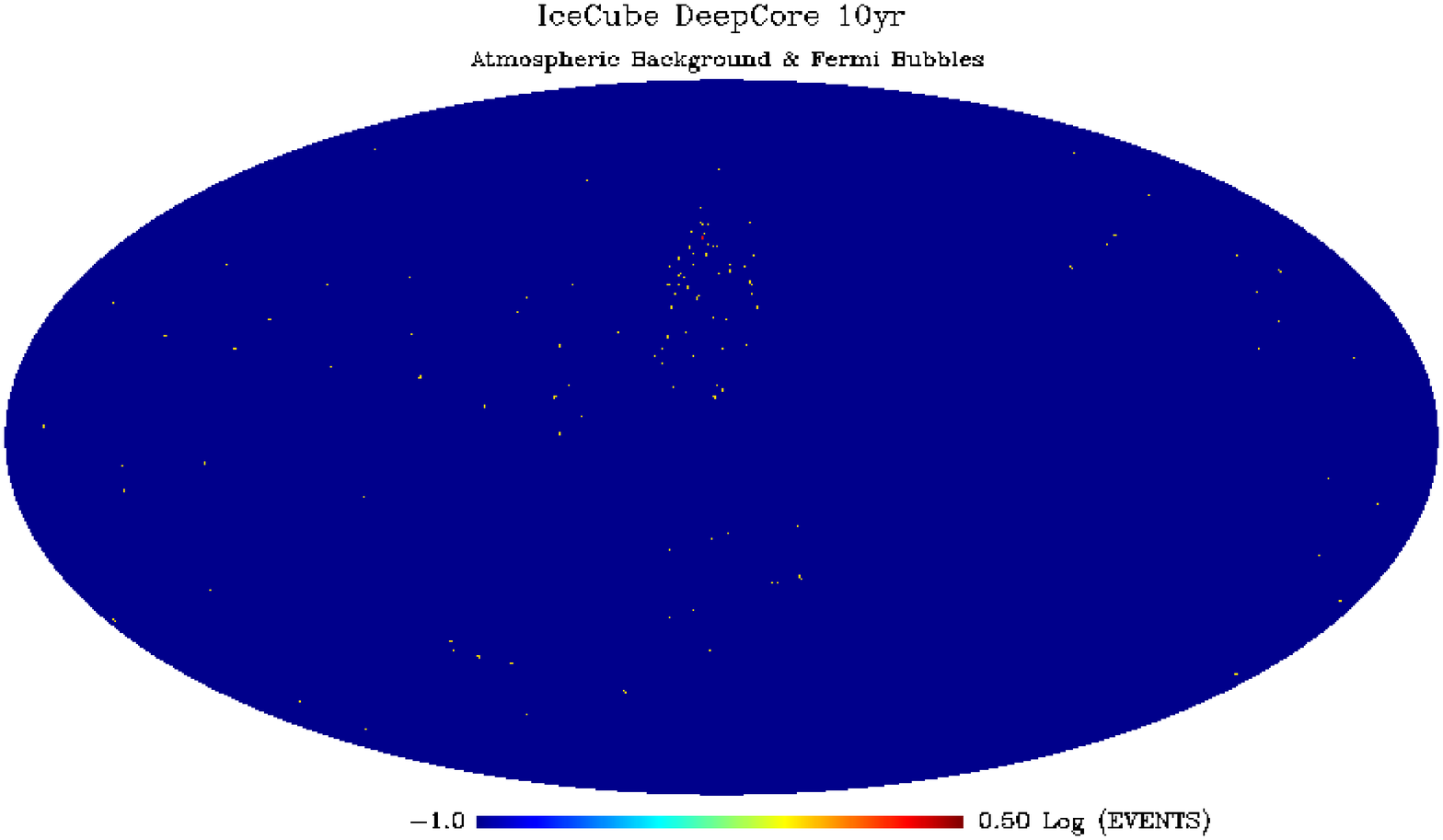}
\hspace{-0.2cm}
\includegraphics[width=3.10in,angle=0]{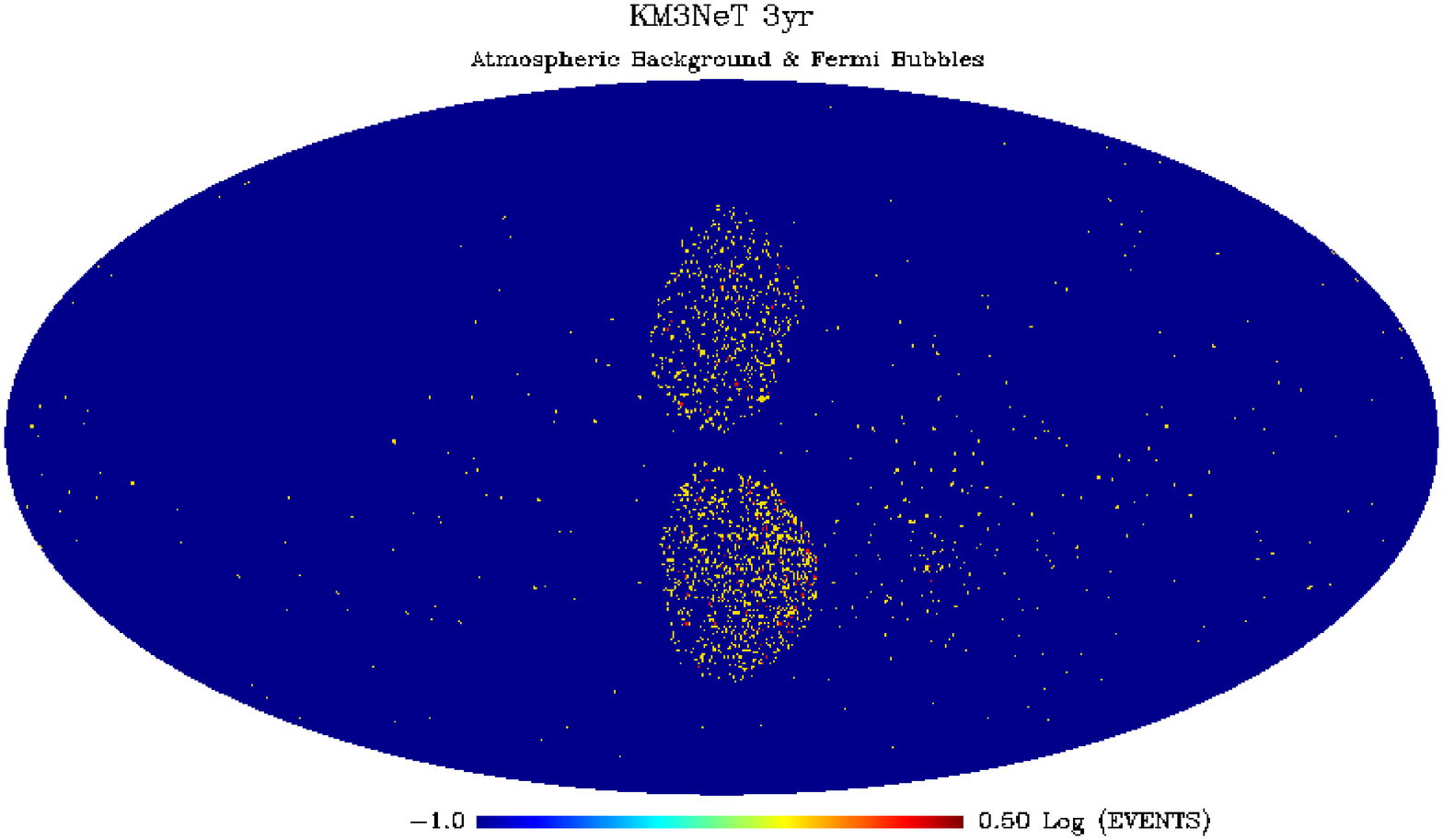}
\caption{$\nu_{\mu} + \overline{\nu}_{\mu}$ events,
with energy between 100 TeV and 1.0 PeV. \textit{Right}: With IceCube 
DeepCore in 10 yr, with "Online Filter" atmospheric
background (63 $\nu_{\mu} + \overline{\nu}_{\mu}$ events) and
contribution from the Fermi Bubbles (74 $\nu_{\mu} + \overline{\nu}_{\mu}$ events). 
\textit{Left}: With KM3NeT in 3yr using HOURS reconstruction technique, 
atmospheric background (465 $\nu_{\mu} + \overline{\nu}_{\mu}$ events) and
contribution from Fermi Bubbles (1795 $\nu_{\mu} + \overline{\nu}_{\mu}$ events).}
\label{fig:BubblesDetect}
\end{figure*}

\begin{figure*}[t]
\includegraphics[width=3.20in,angle=0]{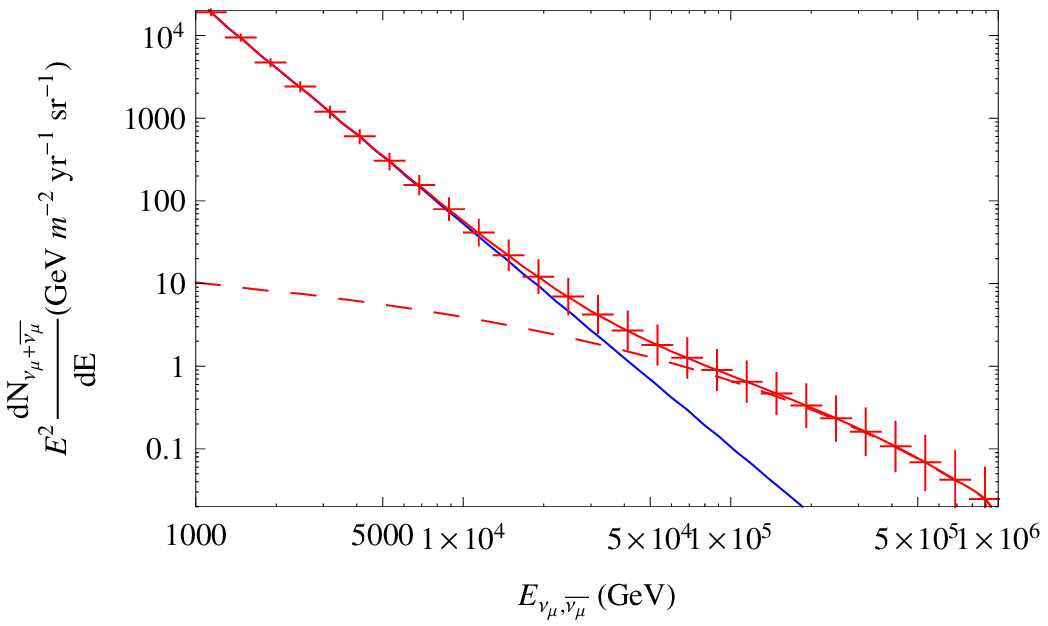}
\hspace{0.1cm}
\includegraphics[width=3.20in,angle=0]{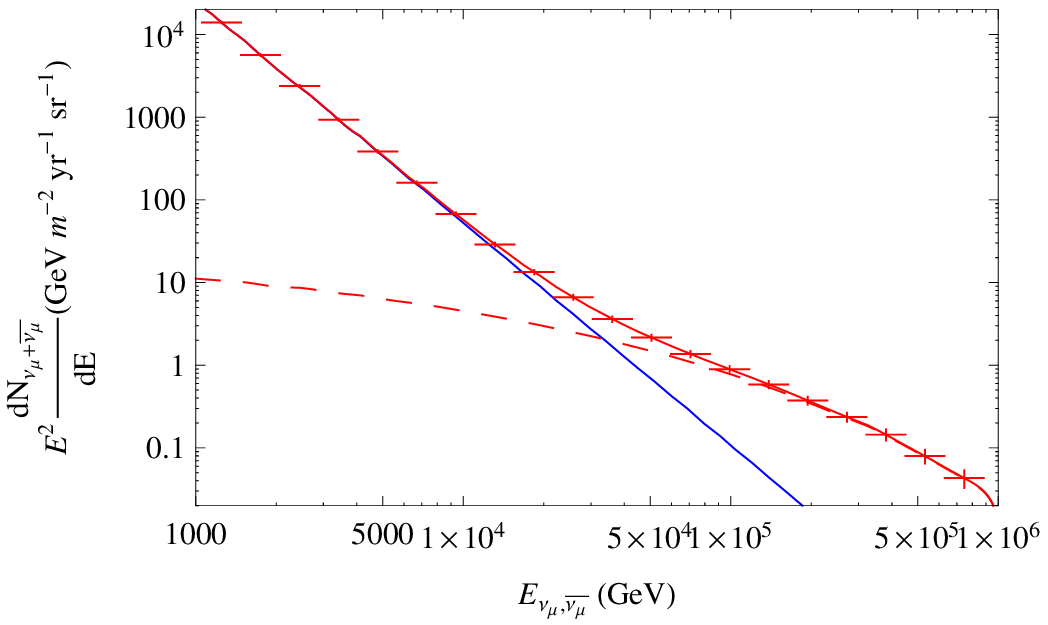}
\caption{$\nu_{\mu} + \overline{\nu}_{\mu}$ reconstructed spectra from Fermi Bubbles. 
\textit{Blue solid line}: atmospheric
background flux, \textit{red bashed lines}: flux from Bubbles,
\textit{red solid lines}: atmospheric+Bubbles flux. \textit{Right}:
IceCube DeepCore with "Online filter" after 3 years of data in the window
of $10^{\circ} < b < 50^{\circ}$ and $0^{\circ} < l < 20^{\circ}$.
\textit{Left}: KM3NeT with "HOURS" reconstruction after 3 years of data
in the window of $-50^{\circ} < b < -10^{\circ}$ and $-20^{\circ} < l < 0^{\circ}$.}
\label{fig:Bubbles3yrReconSpectTel}
\end{figure*}

In Fig.~\ref{fig:BubblesDetect} (left) we show the expected
$\nu_{\mu} + \overline{\nu}_{\mu}$ events mock map, with
energy between 100 TeV to 1 PeV from IceCube DeepCore after 10
yrs of data collection. We show 74 events from the Bubbles
and 63 from the atmospheric background. In Fig.~\ref{fig:BubblesDetect}
(right) we also show what KM3NeT experiment with 3 yrs of reconstructed
events would observe. 

For the Bubbles explanation of \cite{Crocker:2010dg},
masking out or not the disk, can not affect the arguments on detecting
the signal since most of the signal neutrino events are significantly
above or bellow the disk.

Extragalactic point sources also can lead to an additional isotropic 
neutrino component at very high energies where the atmospheric background gets 
suppressed. 
Again the same process (p-p collisions) producing high energy ($\sim$ PeV)
neutrinos will also produce $\gamma$-rays with a spectrum that extends to lower
energies.
Due to the high sensitivity of the \textit{Fermi} LAT
instrument at the 0.1-100 GeV range, the more likely extragalactic sources 
to produce neutrinos have already been detected as point sources in 
$\gamma$-rays as in \cite{Collaboration:2011bm}.
With KM3NeT angular resolution of $> 0.1^{\circ}$ above PeV energies 
it will be very easy to associate even single neutrino events 
to known $\gamma$-ray point sources where p-p collisions is expected to 
be the dominant mechanism for their production (as for instance in star-forming 
galaxies). Alternatively on can mask out known $\gamma$-ray point sources.
Thus extragalactic point sources contribution to the neutrino background can 
be accounted for. Finally cosmogenic neutrinos from Ultra High energy CR (UHECR)
protons interacting with the CMB are already being constrained from the equivalent 
$\gamma$-ray spectrum at the \textit{Fermi} LAT energies 
\cite{Berezinsky:2010xa, Ahlers:2010fw, Decerprit:2011qe, Murase:2012df}
and are expected to be a significant component only at energies above the 
$O(1)$ PeV range. Thus cosmogenic neutrinos do not contribute in the maps of 
Fig.~\ref{fig:BubblesDetect}, that show energies of neutrinos with 
$100$ TeV $< E_{\nu_{\mu},\overline{\nu}_{\mu}} < 1$ PeV 
\footnote{For IceCube DeepCore an $O(1)$ neutrino event per year is estimated
for proton composition of UHECRs \cite{Ahlers:2010fw}}. 

The optimal region for IceCube DeepCore to search for a signal of Bubbles at neutrinos
 would be the left part of the north bubble $10^{\circ} < b < 50^{\circ}$,
$0^{\circ} < l < 20^{\circ}$, for which in Fig.~\ref{fig:Bubbles3yrReconSpectTel}
(left) we show the reconstructed fluxes from atmospheric and from Bubbles after 3 yrs.
Just based on that, one should expect soon a detection of the Bubbles
with IceCube DeepCore if the scenario of \cite{Crocker:2010dg} is correct,
or alternatively setting constraints on the hard CR proton component inside
the Bubbles.

With KM3NeT after 3 yrs as we show in Fig.~\ref{fig:BubblesDetect}
(right) the optimal region is the part of the south bubble with  $-50^{\circ} < b < -10^{\circ}$,
$-20^{\circ} < l < 0^{\circ}$. A clear observation of the morphology (i.e. the 
south and right edge) would be expected. In Fig.~\ref{fig:Bubbles3yrReconSpectTel}
(right) we also show the reconstructed fluxes, which when/if observed at that level, 
would provide a very good measurement also of the injected energy to neutrinos
 (providing an alternative estimate of the CR protons energy).
Alternatively, lack of detection at KM3NeT should exclude the model of 
\cite{Crocker:2010dg} as a possibility for the \textit{Fermi} Bubbles (see also Table~\ref{tab:Sensitivities}).

                                                                                           
\section{AGN and Alternate explanations for the \textit{Fermi} Bubbles}
\label{sec:alternate_models}
 
An alternative explanation for the $\textit{Fermi}$ Bubbles/haze to
those of sections~\ref{sec:LDMA} and~\ref{sec:HSFB} is strong AGN jet 
activity in the Galaxy as \cite{Guo:2011eg, Guo:2011ip}. 
The AGN case is supported by very recent evidence of $\gamma$-ray jets
extending out from the galactic center, together with a $15^{\circ}$ 
width and $40^{\circ}-45^{\circ}$ length cocoon 
at the south galactic hemisphere \cite{Su:2012gu}\cite{DougPrivateCom}.
For the AGN scenario, 
the $\gamma$-ray signal is mainly from CR electrons with energies up to at 
least $\sim$ TeV that up-scatter the local radiation field (mainly CMB at high
latitudes). CR protons can not contribute much of the observed $\gamma$-ray
signal since the ISM gas targets have a very low density at high distances 
(up to 10 kpc) above the disk. Given the CR energy density profiles from 
the MHD simulations of \cite{Guo:2011eg}, it will be very difficult for 
CR protons to explain the morphology of the Bubbles that is relatively flat in latitude,
 suggesting a limb brightening \cite{Mertsch:2011es}. Thus
the scenario of \cite{Guo:2011eg} may give only few neutrino events at 
high latitudes. 

In the 1st \cite{Cheng:2011xd} and 2nd \cite{Mertsch:2011es} order Fermi 
acceleration scenarios, also some protons would be accelerated at high 
energies with power-law spectral indices (for the differential spectrum)
of $E^{-2}$ and $E^{-1}$ \cite{Mertsch:2011es} respectively. Since those protons
would not loose their energy fast (compared to the CR electrons), the
CR proton spectra would be homogeneous within the bubbles. Yet, in both
scenarios the explanation of the $\gamma$-ray signal is entirely from the 
leptonic components.  The protons (at least in those basic scenarios)
are not expected to contribute much to the $\gamma$-ray signal; especially
since in order for protons to contribute significantly to the Bubbles 
signal much greater amounts of total energy to accelerated CRs are needed
than in the models of \cite{Mertsch:2011es, Cheng:2011xd}
\footnote{Also since for CR protons to give $\gamma$-rays p-p collisions
are necessary, the morphology of the $\gamma$-ray signal from these
accelerated protons is not expected to be similar as that of the ICS
$\gamma$-rays from the accelerated electrons.}.
As an example, in the mechanism presented in \cite{Mertsch:2011es}, the 
needed total energy in CR electrons above 100 MeV is $\sim 10^{51}$ erg,
while for the hadronic model of \cite{Crocker:2010dg} the total energy 
in CR protons $\sim 10^{56}$ erg. 
Thus we do not expect a significant number of neutrinos from the leptonic
models of \cite{Mertsch:2011es, Cheng:2011xd} either.

 
\section{Conclusions}
\label{sec:conclusions}

The recent uncovering of the \textit{Fermi} Bubbles/haze 
\cite{Dobler:2009xz, Su:2010qj} in the \textit{Fermi} $\gamma$-ray data 
has generated theoretical work in explaining such a signal in combination (or not) 
with the \textit{WMAP} haze signal of \cite{Finkbeiner:2003im, Dobler:2007wv}. 

We have shown that for the DM explanation of the combined
\textit{Fermi} haze and \textit{WMAP} haze as in \cite{Dobler:2011mk} under 
the annihilation channel to muons that is optimal for neutrinos (XDM \cite{Finkbeiner:2007kk} 
annihilation to $\mu^{\pm}$) and for a prolate DM 
halo we can observe a counterpart signal with a km$^{3}$ telescope located
in the north hemisphere at $\sim$3 yrs of data collection (see discussion in 
section~\ref{sec:LDMA} and Figs.~\ref{fig:KM3NeTMaps3yrRecon}
-~\ref{fig:KM3NeT3yrReconSpect}). 
IceCube DeepCore and ANTARES will not observe any signal for such a 
model; while for the case of XDM annihilation to $e^{\pm}$ through a very 
light scalar boson $\phi < 2 m_{\mu}$ no neutrinos are produced.

For other channels/models of DM annihilation that produce more neutrinos
either from larger suggested boost factors as is 
$\chi\chi \longrightarrow \mu^{+}\mu^{-}$ for the explanation of the 
\textit{Fermi} $e^{+}+e^{-}$ signal (see for example \cite{Bergstrom:2009fa}), 
or due to large hadronic
branching ratios as is $\chi\chi \longrightarrow W^{+}W^{-}$, the neutrino 
events are enhanced significantly. Some
signal from DM is expected even after excluding the disk region where non atmospheric
backgrounds are concentrated (see Fig.~\ref{fig:DMmuonsDetect}).
Yet neutrinos cannot provide the strongest limits for the hadronic channels
of annihilating DM. For leptonic to muons case the limits can be useful 
though, since they can be more robust than the pretty weak limits from 
$\gamma$s and $\bar{p}$ \cite{Cirelli:2008pk, Donato:2008jk, Cholis:2010xb, Evoli:2011id, Cholis:2012am, Cirelli:2008id}.

For the \textit{Fermi} Bubbles explanation of \cite{Crocker:2010dg}, 
a significant number of $> 10$ TeV neutrinos is estimated and even with 
IceCube DeepCore we should expect detection or limits (see Figs.~\ref{fig:BubblesDetect} 
and~\ref{fig:Bubbles3yrReconSpectTel}).
Furthermore, a km$^{3}$ telescope in the north hemisphere will either exclude the model
of \cite{Crocker:2010dg} or confirm the morphology of the Bubbles at
 neutrinos and measure the injected power to neutrinos from inelastic 
pp collisions inside the Bubbles as we show in Figs.~\ref{fig:BubblesDetect}
and~\ref{fig:Bubbles3yrReconSpectTel}.

Leptonic scenarios for the \textit{Fermi} Bubbles such as those of 
\cite{Guo:2011eg, Mertsch:2011es, Cheng:2011xd} would 
also predict some CR protons, but are not expected to give any significant 
neutrino signal, since the main source for the $\gamma$-rays is IC 
scattering from CR electrons.  
  
With the current IceCube DeepCore telescope a first probe to some of the models 
of the \textit{Fermi} Bubbles/haze can be achieved, while with a km$^{3}$ telescope
located in the north hemisphere, discrimination between the hadronic \cite{Crocker:2010dg},
 the leptonic \cite{Guo:2011eg, Mertsch:2011es, Cheng:2011xd} and the DM 
\cite{Dobler:2011mk} cases will be attained.  

As this paper was being written, a new analysis of \textit{Fermi} 
$\gamma$-rays has suggested the evidence of $\gamma$-rays jets in the 
Milky Way extending out to $\sim 10$ kpc from the galactic center 
\cite{Su:2012gu}. Additionally, a cocoon structure has been revealed in the 
southern galactic hemisphere. While according to \cite{Su:2012gu} the total luminosity 
of the north and the south jet-like features is $(1.8 \pm 0.35)\times 10^{35}$ 
erg/s at 1-100 GeV i.e. 2 orders of magnitude less in luminosity than the 
Bubbles ($2\times10^{37}$ erg/s in the same energy range), such an additional 
signal favors the AGN case, with the Bubbles coming from the decelerated 
jet material \cite{Su:2012gu}.

Yet the two signals, i.e the combined $\gamma$-ray cocoon and jets and the 
\textit{Fermi} Bubbles may be created at a different time. Such a case
would allow for a combination of sources, DM $\&$ AGN, Hadronic model $\&$ AGN
to account for the total $\gamma$-ray Bubbles/haze $\&$ jets and cocoon signals.
For the hybrid scenario of DM $\&$ AGN, the edge at $l > 0$ (left) of the south bubble  
 could just be the result of the presence (overlap on the sky) of 
the AGN cocoon and thus morphologically would not need to be explained by DM annihilation. 
Similarly a cocoon in the northern galactic hemisphere 
(not claimed to be clearly revealed yet) could account for the north to the right 
edge (at $l < 0$) in the data. Additionally the AGN responsible for the jets and the cocoon of 
\cite{Su:2012gu} could evacuate the cavity of the Bubbles which then the high
energy $e^{\pm}$ from annihilating DM fill \cite{Dobler:2011mk}.        

Since the luminosity of the jets alone is only 1$\%$ of that of the
Bubbles/haze, assuming that the current state of the jets is representative of the 
time-averaged state\footnote{As noted by D. Finkbeiner if currently the 
jets are at a low luminosity phase, then the time averaged luminosity of the jets could be 
much higher\cite{DougPrivateCom}. In such a case the entire Bubbles signal could come from 
 AGN activity.}, the predictions for the neutrino fluxes from the DM/Hadronic
cases remain the same\footnote{The spectrum of the cocoon revealed only in 
the south galactic hemisphere, 
is similar to that of the jets and the Bubbles \cite{Su:2012gu} but still 
covers only part of the south Bubble angular region thus it can not account 
for the major part of the south Bubble $\gamma$-ray luminosity.}.

\vskip 0.2 in
\section*{Acknowledgments}
The author would like to thank Gregory Dobler, Douglas Finkbeiner, Maryam Tavakoli, Aikaterini Tzamarioudaki, 
Spyros Tzamarias, Piero Ullio and Neal Weiner for valuable discussions. The author 
would also like to specifically thank Apostolos Tsirigotis for offering
information regarding the HOURS simulation of events reconstruction in KM3NeT.    

\vskip 0.05in


\bibliography{NeutrinoHaze}
\bibliographystyle{apsrev}

\end{document}